\documentclass[%
reprint,
amsmath,amssymb,
aps,]{revtex4-1} 

\usepackage{graphicx}% Include figure files
\usepackage{dcolumn}% Align table columns on decimal point
\usepackage{bm}% bold math
\usepackage{xcolor}

\def\l{\langle}
\def\r{\rangle}

\begin{document}
\title{
Large peaks in the entropy of the diluted nearest-neighbor spin-ice model\\
on the pyrochlore lattice in a [111] magnetic field
} 

\author{Petr Andriushchenko$^{1,2}$}
\email{pitandmind@gmail.com}
\author{Konstantin Soldatov$^{1,2}$}
\email{soldatov\_ks@students.dvfu.ru}
\author{Alexey Peretyatko$^{1}$}
\email{peretiatko.aa@dvfu.ru}
\author{Yuriy Shevchenko$^{1,2}$}
\email{shevchenko.ya@dvfu.ru}
\author{Konstantin Nefedev$^{1,2}$}
\email{nefedev.kv@dvfu.ru}
\author{Hiromi Otsuka$^{3}$}
\email{otsuka@phys.se.tmu.ac.jp}
\author{Yutaka Okabe$^{3}$}
\email{okabe@phys.se.tmu.ac.jp}
\affiliation{
$^1$School of Natural Sciences, Far Eastern Federal University, Vladivostok, 8 Sukhanova, 690091, Russian Federation \\
$^2$Institute of Applied Mathematics, Far Eastern Branch, Russian Academy of Science, Vladivostok, 7 Radio, 690041, Russian Federation \\
$^3$Department of Physics, Tokyo Metropolitan University, Hachioji, Tokyo 192-0397, Japan \\
}

\date{\today}

\begin{abstract}
We study the residual entropy of the nearest-neighbor spin-ice model 
in a magnetic field along the [111] direction using the Wang-Landau 
Monte Carlo method, with a special attention to dilution effects.
For a diluted model, we observe a stepwise decrease 
of the residual entropy as a function of the magnetic field, 
which is consistent with the finding of the five magnetization plateaus 
in a previous replica-exchange Monte Carlo study 
by Peretyatko {\it et al.} [Phys. Rev. B {\bf 95}, 144410 (2017)]. 
We find large peaks of the residual entropy due to the degeneracy 
at the crossover magnetic fields, $h_c/J$ = 0, 3, 6, 9, and 12, 
where $h$ and $J$ are the magnetic field and 
the exchange coupling, respectively. 
In addition, we also study the residual entropy of 
the diluted antiferromagnetic 
Ising models in a magnetic field on the kagome and triangular lattices. 
We again observe large peaks of the residual entropy, 
which are associated with multiple magnetization plateaus 
for the diluted model.  
Finally, we discuss the interplay of dilution and magnetic fields 
in terms of the residual entropy.
\end{abstract}


\maketitle

\section{Introduction}

Frustrated spin systems have recently drawn considerable interest. 
The discovery of the spin-ice compounds, such as Dy${_2}$Ti${_2}$O${_7}$ 
and Ho${_2}$Ti${_2}$O${_7}$ on a pyrochlore lattice, 
has accelerated studies into the mechanism related to 
frustration \cite{Harris,Ramirez}.  
The existence of the residual macroscopic entropy is one of 
the areas of interest in frustrated systems, which was first discussed 
by Pauling for water ice~\cite{Pauling}. 
In the spin-ice materials, the magnetic ions (Dy$^{3+}$ 
or Ho$^{3+}$) occupy a pyrochlore lattice formed by 
corner-sharing tetrahedra. The local crystal field environment 
aligns the magnetic moments in the directions 
connecting the centers of two tetrahedra at low 
temperatures~\cite{Bramwell,Diep}. In the low-temperature spin-ice state, 
the magnetic moments are highly constrained locally and obey 
the so-called ``ice rules" as in a water ice, that is, two spins point in 
and two spins point out of each tetrahedron of the pyrochlore lattice. 

Magnetic field effects, especially the existence of magnetization plateaus, 
have been studied theoretically \cite{Harris98,Moessner,Isakov04} 
and experimentally 
\cite{Matsuhira02,Hiroi,Sakakibara,Higashinaka,Fukazawa}. 
Isakov {\it et al.}~\cite{Isakov04} found a large peak 
in the entropy between the two plateaus.  

The dilution effect on frustration is another topic of spin-ice 
materials, and Ke {\it et al.}~\cite{Ke} studied the dilution 
effects by replacing the magnetic ions Dy$^{3+}$ or Ho$^{3+}$ 
by nonmagnetic Y$^{3+}$ ions. 
Nonmonotonic zero-point entropy as a function of dilution concentration 
was observed experimentally, and a generalization of Pauling's 
theory of the residual entropy was discussed \cite{Ke}. 
Detailed experimental studies combined 
with Monte Carlo simulations have been reported~\cite{Lin,Scharffe}. 

Recently, Peretyatko {\it et al.}~\cite{Peretyatko}
studied the effect of magnetic fields on diluted spin-ice models 
in order to elucidate the interplay of dilution and magnetic fields. 
Five plateaus were observed in the magnetization curve of the diluted 
nearest-neighbor (NN) spin-ice model on the pyrochlore lattice 
when a magnetic field was applied in the [111] direction. 
This effect is in contrast with the case of a pure (i.e.,~undiluted) model, 
which displays two plateaus. 

In this paper, we present the diluted spin-ice model 
on the pyrochlore lattice, focusing on the entropy, 
when a magnetic field is applied in the [111] direction. 
In order to investigate the entropy, 
we use the Wang-Landau (WL) Monte Carlo method~\cite{WL}, 
which directly calculates the energy density of states (DOS), 
$g(E)$.  The precise estimates of the residual entropy for 
the diluted spin-ice model with no magnetic field were 
reported by Shevchenko {\it et al.}~\cite{Shevchenko}.
If we use a canonical Monte Carlo simulation such as 
the Metropolis algorithm, the estimate of the entropy is obtained 
by the numerical integration of the specific heat.  
Instead, using the WL method, we can compute the entropy 
in a straightforward way. 

As a theoretical model of the spin-ice material, in this study, 
we treat the NN antiferromagnetic (AFM) Ising model 
on the pyrochlore lattice. 
A more complicated model, such as the dipolar model, 
may be required to make connections to actual materials.
However, Isakov {\it et al.} \cite{Isakov05} discussed 
the reason for which the low-temperature entropy of the spin-ice 
compounds is well described by the NN AFM Ising model 
on the pyrochlore lattice, i.e., by the "ice rules". 

In this paper, we also deal with the diluted AFM Ising model 
on the kagome and triangular lattices as other frustrated systems. 
We study the magnetic-field dependence of the residual entropy 
for the diluted model. The study of the kagome lattice 
is instructive for the comparison with the model 
on the pyroclore lattice.  The pyrochlore lattice can be regarded 
as alternating kagome and triangular layers, 
and the magnetic field in the [111] direction effectively 
decouples these layers. The spins in the triangular layers 
are fixed when the magnetic field is applied in this direction. 
The behavior of the spins in the kagome layers is therefore 
of significant interest, and is sometimes referred to 
as the "kagome-ice" problem. Thus, the comparison 
of the dilution effects of the magnetization curve and the 
residual entropy between the "kagome-ice" state in the pyrochlore 
lattice and the two-dimensional (2D) kagome lattice is interesting.

This paper is organized as follows: 
Section II and III describe the model and the simulation method, 
respectively. The results are presented 
and discussed in Section IV. Section V is devoted to 
the summary and discussions.  
A review of the theory of the pure models is 
presented in the Appendix.
There, we give the exact numerical estimates of 
the magnetization and entropy up to 16 digits 
at the crossover field 
for the pure AFM Ising model on the triangular lattice 
in a magnetic field.

%%%%%%%%%%%%%%%%%%%%%%%%%%%%%%%%%%%%%%%%%%%%%%%%%%%%%%%%%%%%%%%%%%%%%%%%%%%%
\begin{figure}
\begin{center}
\includegraphics[height=6.0cm]{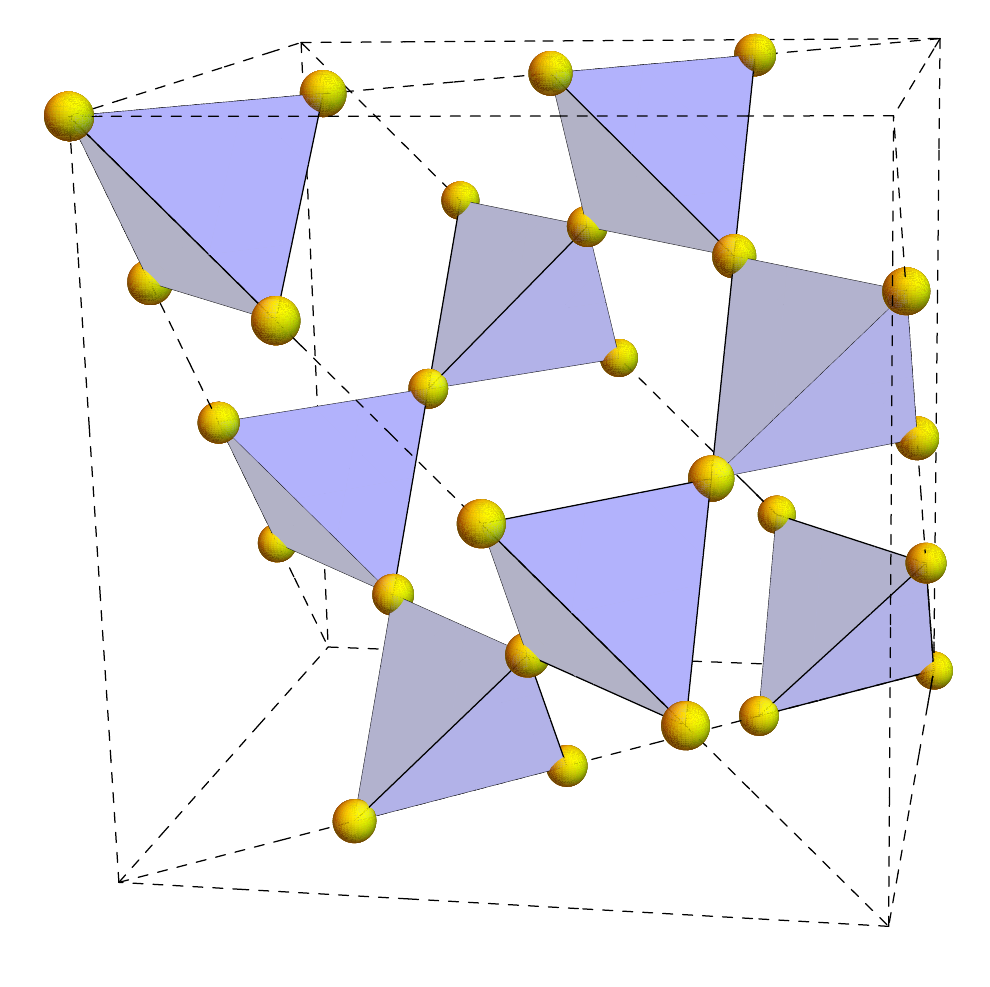}
\caption{
Illustration of the 16-site cubic unit cell 
of the pyrochlore lattice.
}
\label{fig:pyro_illust}
\end{center}
\end{figure}
%%%%%%%%%%%%%%%%%%%%%%%%%%%%%%%%%%%%%%%%%%%%%%%%%%%%%%%%%%%%%%%%%%%%%%%%%%%%

\section{Model}

We investigate the AFM Ising model with NN interaction 
on the pyrochlore lattice, which is displayed in  Fig.~\ref{fig:pyro_illust}. 
For the simulation, we use the 16-site cubic unit cell of 
the pyrochlore lattice. The systems with $L \times L \times L$ 
unit cells have $N=16 * L^3$ sites. 
When a magnetic field $\bm{h}$ is applied in the spin-ice model, 
the Hamiltonian is given by~\cite{Isakov04}
\begin{equation}
 H = J \sum_{\l i,j \r} \sigma_i \sigma_j
   - \sum_i \bm{h} \cdot \bm{d}_{\kappa(i)} \ \sigma_i, 
\label{Hamiltonian}
\end{equation}
where $J (>0)$ is the effective AFM coupling, 
$\sigma_i$ are the Ising pseudo-spins ($\sigma_i = \pm 1$), and 
$\l i,j \r$ stands for the NN pairs. 
The unit vectors $\bm{d}_{\kappa(i)}$ represent the local easy axes 
of the pyrochlore lattice, and explicitly described as 
$
\bm{d}_{\kappa(i)} = \{ \bm{d}_0, \bm{d}_1, \bm{d}_2, \bm{d}_3 \}, 
$
where
$\bm{d}_0 = (1,1,1)/\sqrt{3}$, 
$\bm{d}_1 = (1,-1,-1)/\sqrt{3}$, 
$\bm{d}_2 = (-1,1,-1)/\sqrt{3}$, 
and 
$\bm{d}_3 = (-1,-1,1)/\sqrt{3}$.
When the magnetic field $\bm{h}$ is along 
the [111] direction, that is,
$
   \bm{h} = h \bm{d}_0,
$
$\bm{h} \cdot \bm{d}_{\kappa(i)}$ becomes $h$ 
for apical spins where $\bm{d}_{\kappa(i)} = \bm{d}_0$, 
but $-(1/3)h$ for other spins. 
We calculate the magnetization $M$ along the [111] direction 
using the relation
\begin{equation}
   M = \sum_i \bm{d}_0 \cdot \bm{d}_{\kappa(i)} \ \sigma_i.
\label{magnetization}
\end{equation}

In the case of the site dilution of spins, the Hamiltonian becomes
\begin{equation}
 H = J \sum_{\l i,j \r} c_i c_j \sigma_i \sigma_j
   - \sum_i \bm{h} \cdot \bm{d}_{\kappa(i)} \ c_i \sigma_i, 
\label{Hamiltonian_dilute}
\end{equation}
where $c_i$ represent the quenched variables ($c_i = 1 \ {\rm or} \ 0$), 
and the concentration of vacancies is denoted by $x$. Then, 
the magnetization becomes
\begin{equation}
   M = \sum_i \bm{d}_0 \cdot \bm{d}_{\kappa(i)} \ c_i \sigma_i.
\label{magnetization_dilute}
\end{equation}
%

%%%%%%%%%%%%%%%%%%%%%%%%%%%%%%%%%%%%%%%%%%%%%%%%%%%%%%%%%%%%%%%%%%%%%%%%%%%%
\begin{figure}
\begin{center}
\includegraphics[width=3.6cm]{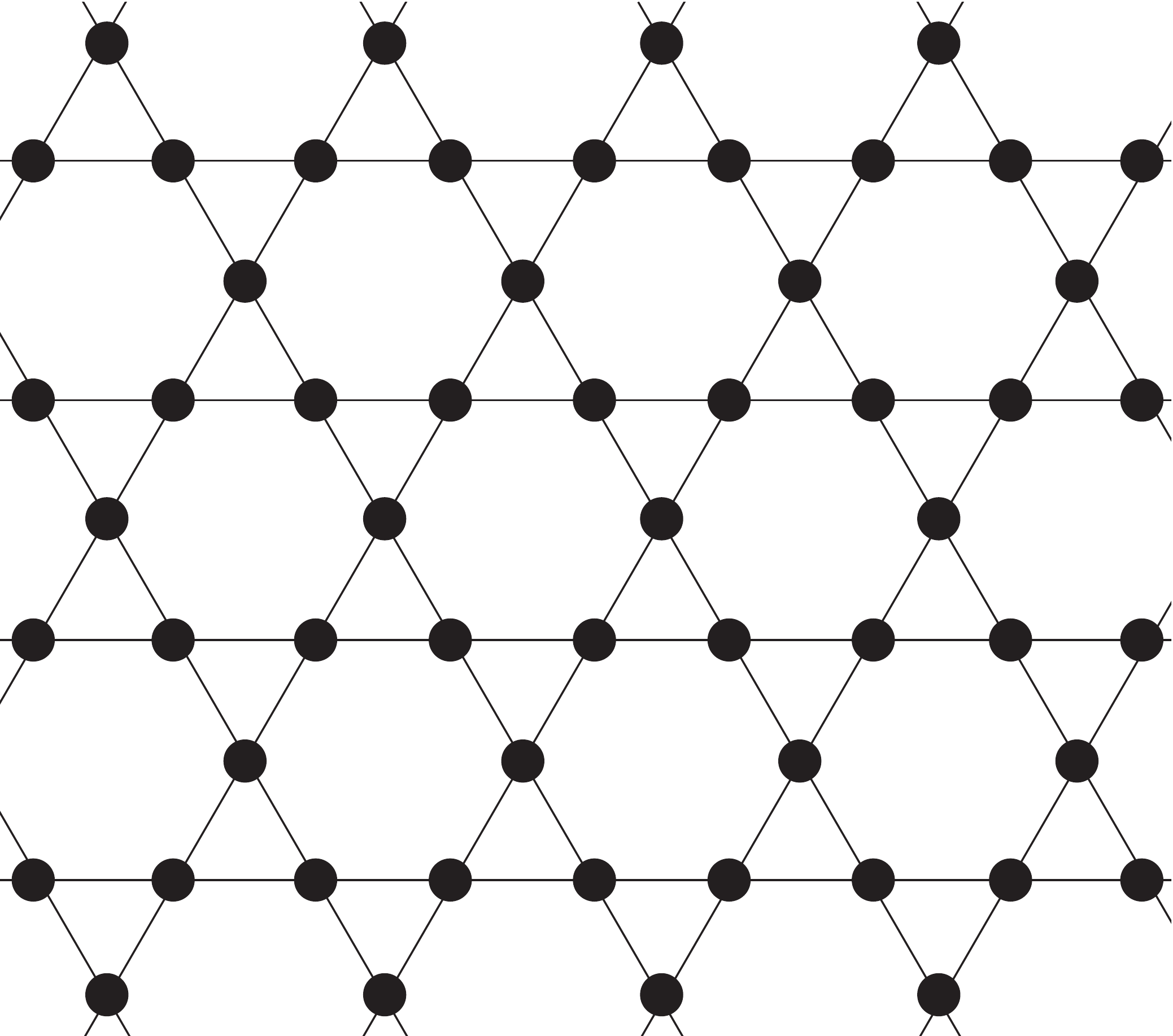}
\hspace{0.4cm}
\includegraphics[width=4.0cm]{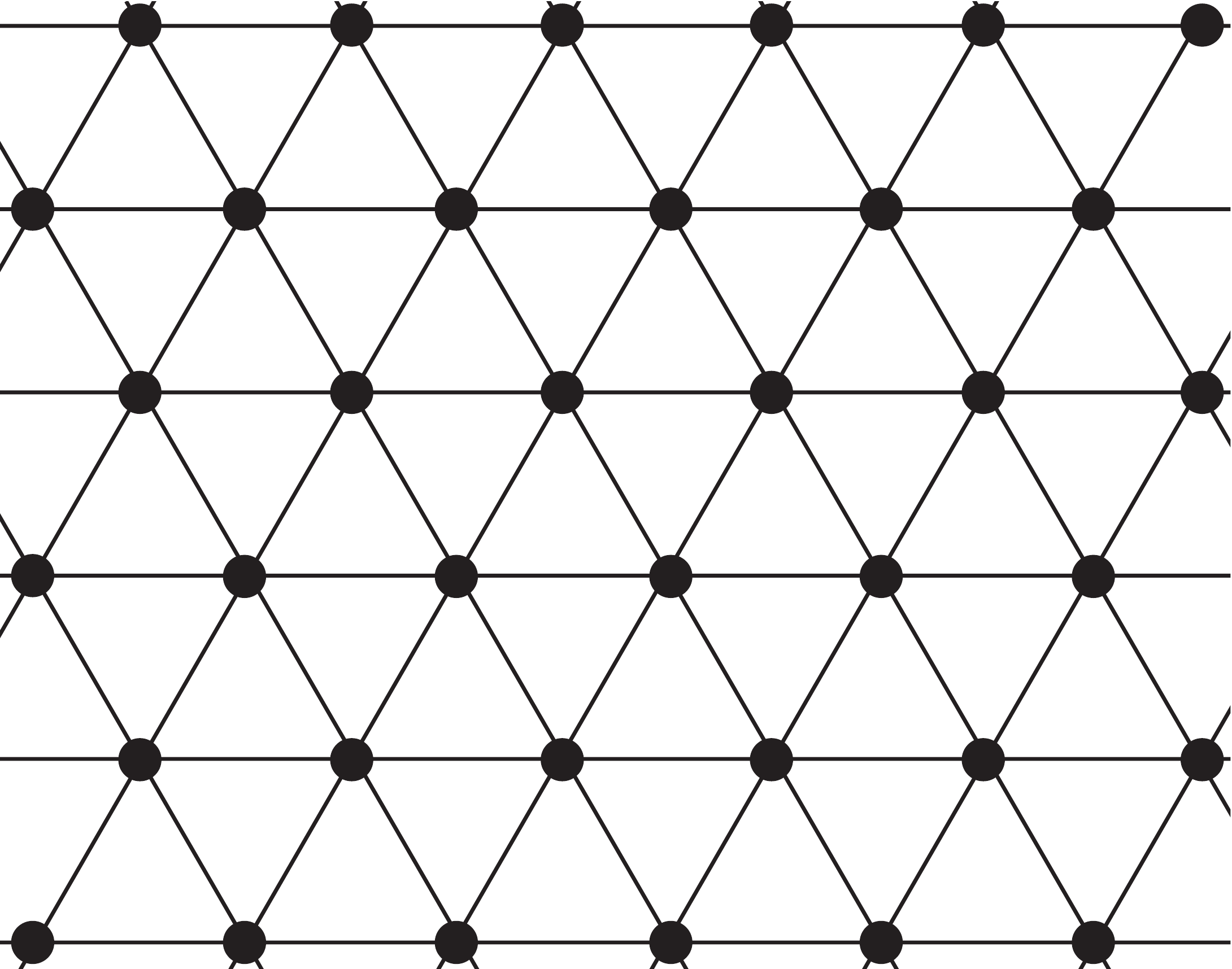}
\caption{
Illustration of the kagome (left) and triangular (right) lattices.
}
\label{fig:kagome_tri_illust}
\end{center}
\end{figure}
%%%%%%%%%%%%%%%%%%%%%%%%%%%%%%%%%%%%%%%%%%%%%%%%%%%%%%%%%%%%%%%%%%%%%%%%%%%%

In addition, we treat the diluted AFM Ising models on the kagome and triangular 
lattices; these lattices are illustrated in Fig.~\ref{fig:kagome_tri_illust}. 
In these models, the Zeeman term of the Hamiltonian is simply 
given by $-hM$, where the magnetization is calculated by
\begin{equation}
  M = \sum_i c_i \sigma_i.
\end{equation}

\section{Simulation Method}

We used the complete enumeration over all possible configurations of system,
which is the most straightforward but computationally expensive way 
to obtain the information of the system entropy.

For bigger systems we used the WL method that directly calculates the energy DOS 
to obtain precise numerical information on the entropy of the system with hight accuracy. 
To treat the effects of an external field in the WL method, 
one may consider two types of approaches. 
On one hand, the Zeeman term, which is the second term 
in Eqs.~(\ref{Hamiltonian}) and (\ref{Hamiltonian_dilute}), 
is directly included in the calculation of the energy DOS.  
On the other hand, the joint DOS $g(E_0,M)$ of the Hamiltonian 
without the Zeeman term for two variables, the exchange energy $E_0$ 
and the magnetization $M$, can be calculated. 
The single DOS $g(E)$ can be calculated using the constraint of $E=E_0-hM$. 
The latter approach was employed for a pure model 
by Ferreyra {\it et al.}~\cite{Ferreyra}. 
However, this approach has a disadvantage 
in that it is too computationally intensive, and the calculations 
are therefore limited to smaller system sizes. 
In Ref.~\cite{Ferreyra}, the authors treated up to $L=3$ ($N=432$). 
Here, we used the direct approach of calculating 
the DOS of the Hamiltonian including the Zeeman term, 
and the system sizes increase up to $L=5$ ($N=2000$), and up to 
$L=6$ ($N=3456$) for a pure model.

In the WL algorithm, a random walk in energy space is performed 
with a probability proportional to the reciprocal of the DOS, 
$1/g(E$), which results in a flat histogram of energy distribution. 
We make a move based on the transition probability 
from energy level $E_1$ to $E_2$:
\begin{equation}
  p(E_1 \to E_2) = \min\Big[1, \frac{g(E_1)}{g(E_2)}\Big].
\end{equation}
Since the exact form of $g(E)$ is not known {\it a priori}, 
we determine $g(E)$ iteratively by introducing the modification 
factor $f_i$.  Then, every time the state is visited, 
$g(E)$ is modified by 
\begin{equation}
  \ln g(E) \to \ln g(E) + \ln f_i.
\label{g_update}
\end{equation}
At the same time the energy histogram $h(E)$ is updated as
\begin{equation}
  h(E) \to h(E) + 1.
\label{h_update}
\end{equation}
The modification factor $f_i$ is gradually reduced to unity by
checking the ``flatness" of the energy histogram. 
Then $f_i$ is modified as
\begin{equation}
  \ln f_{i+1} = \frac{1}{2} \ln f_i,
\end{equation}
and the histogram $h(E)$ is reset. As an initial value of $f_i$, 
we choose $f_0 = e$; as a final value, we choose $\ln f_i = 2^{-22}$, 
that is, $f_{22} \simeq 1.00000024$.

The ratio of $g(E)$ for different energies $E_1$ and $E_2$, 
$g(E_1)/g(E_2)$, can be calculated in the WL algorithm. 
If we are interested only in the temperature dependence of 
the physical quantities, such as the total energy or the specific heat, 
the ratio of $g(E)$ is sufficient. However, to determine the absolute value 
of the entropy, the normalization of $g(E)$ is necessary.  In the case of 
the Ising model, each spin takes one of two states; thus, 
the normalization condition becomes
\begin{equation}
   \sum_E g(E) = 2^{N_{\rm spin}},
\label{normalization}
\end{equation}
where $N_{\rm spin}$ is the number of spins.  In the case of dilution, 
the number of spins $N_{\rm spin}$ is different 
from the number of sites $N$. 

Here, we describe the technical details of the WL method. 
The system with the magnetic field is asymmetric for the inversion of 
the whole spins. Thus, the states with totally different spin 
configurations may have the same energy; 
it takes long time to adjust $g(E)$ for such a case. 
If we separate the states into two parts, one part where 
the total magnetization is positive, and the other where 
the total magnetization is negative, 
the convergence becomes much faster.  
In the system of pyrochlore lattice, we had better use 
the total magnetization of Ising pseudo-spins, 
$$
 M_p = \sum_i \sigma_i,
$$
instead of $M$, Eqs.~(\ref{magnetization}) and 
(\ref{magnetization_dilute}), for this purpose. 
We normalize the DOS as 
\begin{equation}
   \sum_E g_{+, -}(E) = 2^{N_{\rm spin}/2}
\label{normalization_pm}
\end{equation}
for each configuration space. Precisely speaking, we should 
consider the contribution of the number of 
the states that the total magnetization is equal to zero 
for even $N_{\rm spin}$.  However, this contribution 
is negligibly small for large $N_{\rm spin}$. 
When we restrict the space of the random walk to only 
the positive magnetization space or the negative magnetization space, 
we should consider the treatment of the boundary. 
It was argued in the case of the restricted energy interval~\cite{Schulz} 
that when the suggested spin flip is out of range, 
we should reject the suggested spin flip and 
count the current energy level once more 
in Eqs.~(\ref{g_update}) and (\ref{h_update}). 

For a pure system, we have the information on the ground 
state based on theoretical analysis, such that 
the ground state is the two-in two-out configuration 
for a certain range of the magnetic field. 
However, for a diluted case, we do not know 
the ground state. The search for the ground state 
is a difficult problem for random frustrated systems, 
such as spin glasses~\cite{Berg}.  
In this paper, we follow a two-step process for some 
diluted systems where the convergence is very slow. 
First, we search for the ground state by simulating 
the restricted range of energy. Then, as a second step, 
we perform the WL simulation for a full range of energy 
using the information obtained in the first step 
to estimate the absolute value of the entropy 
based on the normalization condition, i.e., 
Eq.~(\ref{normalization_pm}). 

The convergence and refinement of the WL algorithm have been 
discussed by many researchers~\cite{Zhou,Lee,Belardinelli}. 
However, most of the works aim to obtain very good convergence 
of symmetric systems, for example, the pure Ising model. 
Several approaches employed in this paper are also useful 
for other asymmetric or random frustrated systems~\cite{Okabe02}. 

Here, we summarize the conditions 
of our simulation. The system sizes of the pyrochlore lattice 
are $L$ = 3, 4, and 5; the numbers of sites is $N$ = 432, 
1024, and 2000, respectively. For a pure system, 
we also treated $L=6$ ($N=3456$).
The simulation of the AFM Ising model on the kagome lattice 
considered systems of size $L \times (3/2)L$ 
with $L = 36$ ($N = 1944$) and $L = 48$ ($N = 3456$). 
For the triangular lattice of size $L \times L$, we used 
$L = 36$ ($N = 1296$) and $L = 48$ ($N = 2304$). 
In all of the cases we used the periodic boundary conditions. 
With respect to the dilution concentration $x$, we treat 
$x$= 0.0 (pure), 0.2, 0.4, 0.6, and 0.8. 
For each calculation, we took an average of 10 samples. 
For a pure system, this means that we performed simulations 
with different random-number sequences. For a diluted system, 
different random configurations of dilution were treated.

\section{Results}

\subsection{Pure models}

%%%%%%%%%%%%%%%%%%%%%%%%%%%%%%%%%%%%%%%%%%%%%%%%%%%%%%%%%%%%%%%%%%%%%%%%%%%%
\begin{figure}
\begin{center}
\includegraphics[width=8.5cm]{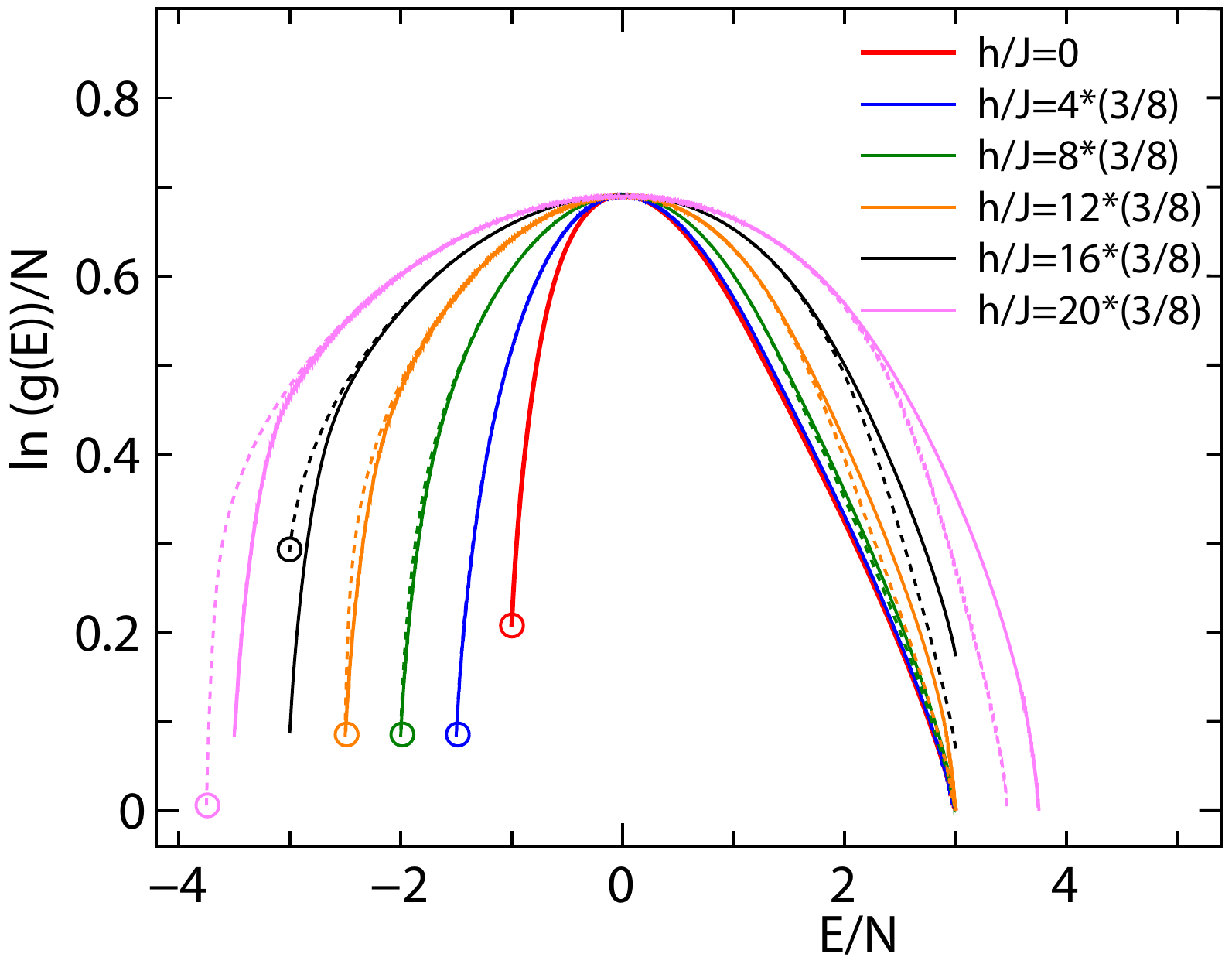}
\caption{
Plot of $\ln g(E)$ as a function of $E$
of the Ising model on the pyrochlore lattice 
for several values of magnetic field $h$. 
The system size is $L=5 \ (N=2000)$. 
The DOS of positive pseudo-spin magnetization configurations,
$g_+(E)$, and that of negative magnetization configurations,
$g_-(E)$, are plotted by dotted and solid curves, 
respectively. The positions of the ground state are encircled 
in the figure.    
}
\label{fig:pyro_DOS}
\end{center}
\end{figure}
%%%%%%%%%%%%%%%%%%%%%%%%%%%%%%%%%%%%%%%%%%%%%%%%%%%%%%%%%%%%%%%%%%%%%%%%%%%%

We first show the results of pure (i.e., undiluted) systems ($x=0$). 
The theory of pure AFM Ising models in a magnetic field 
on frustrated lattices is summarized in the Appendix. 
There, we give the results obtained for the triangular lattice, 
the kagome lattice, and the pyrochlore lattice. 
For the pyrochlore lattice, the magnetic field 
is applied in the [111] direction. 

In the WL simulation, we directly calculate the energy DOS. 
As an example, in Fig.~\ref{fig:pyro_DOS}, we plot $\ln(g(E))/N$, 
essentially the entropy, as a function of $E$ (in units of $J$) 
of the Ising model on the pyrochlore lattice for several values 
of magnetic field $h$. 
The system size is $L=5$ ($N=2000$), 
and the plot is of one sample for each $h$. 
The magnetic field was chosen as a multiple of $(3/8)J$ 
because we are making integer calculation for a refined 
interval of the magnetic field. 
The energy takes a value 
from $-N(J+h/3)$ to $3NJ$ for $h/J<6$ and 
from $-N(h/2)$ to $N(h/2)$ for $h/J>6$. 
The DOS of positive pseudo-spin magnetization configurations,
$g_+(E)$, and that of negative pseudo-spin magnetization configurations,
$g_-(E)$, are plotted by dotted and solid curves, 
respectively. The positions of the ground state are encircled 
in the figure.    

%%%%%%%%%%%%%%%%%%%%%%%%%%%%%%%%%%%%%%%%%%%%%%%%%%%%%%%%%%%%%%%%%%%%%%%%%%%%
\begin{figure*}
\begin{center}
\includegraphics[width=7.0cm]{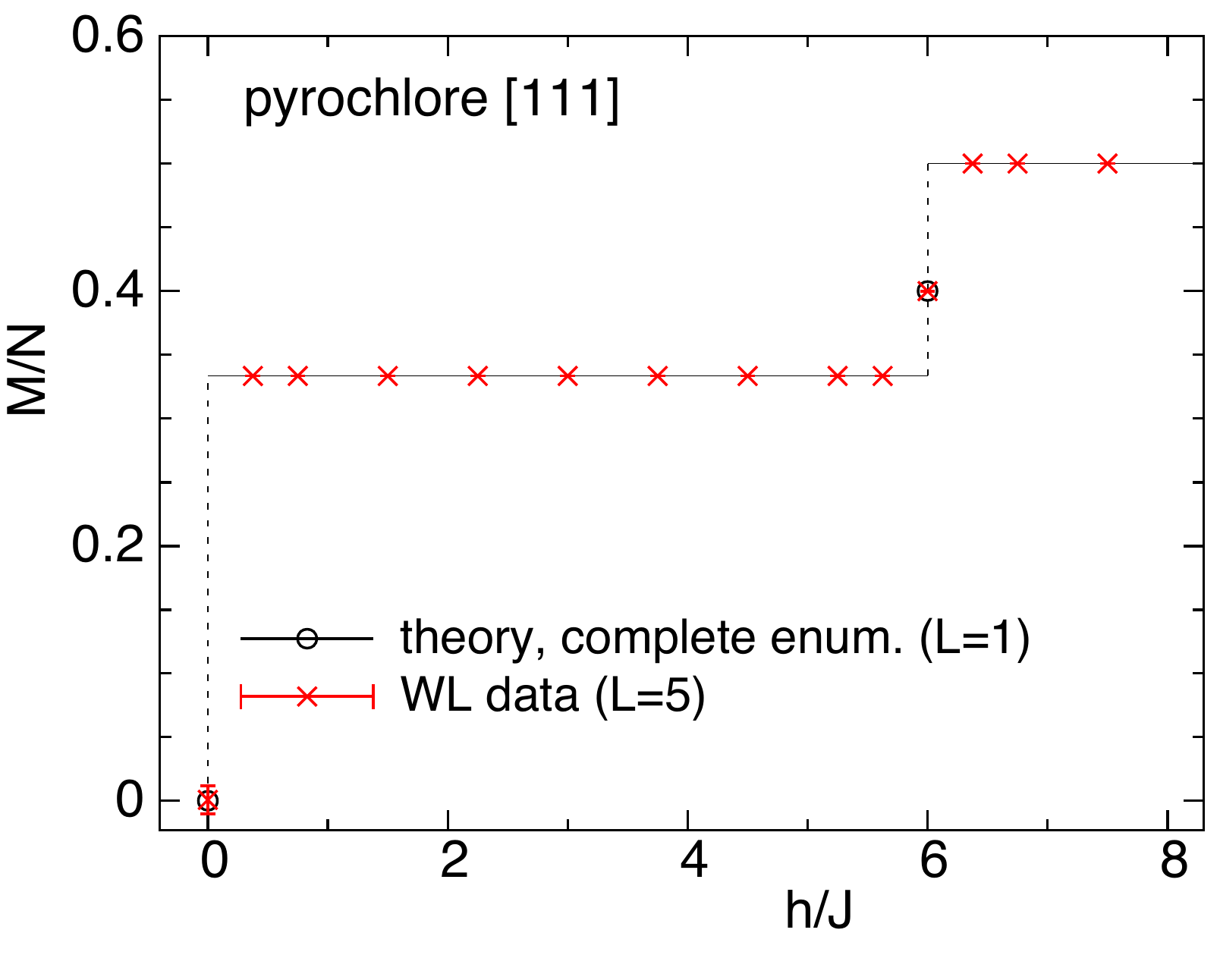}
\hspace{12mm}
\includegraphics[width=7.0cm]{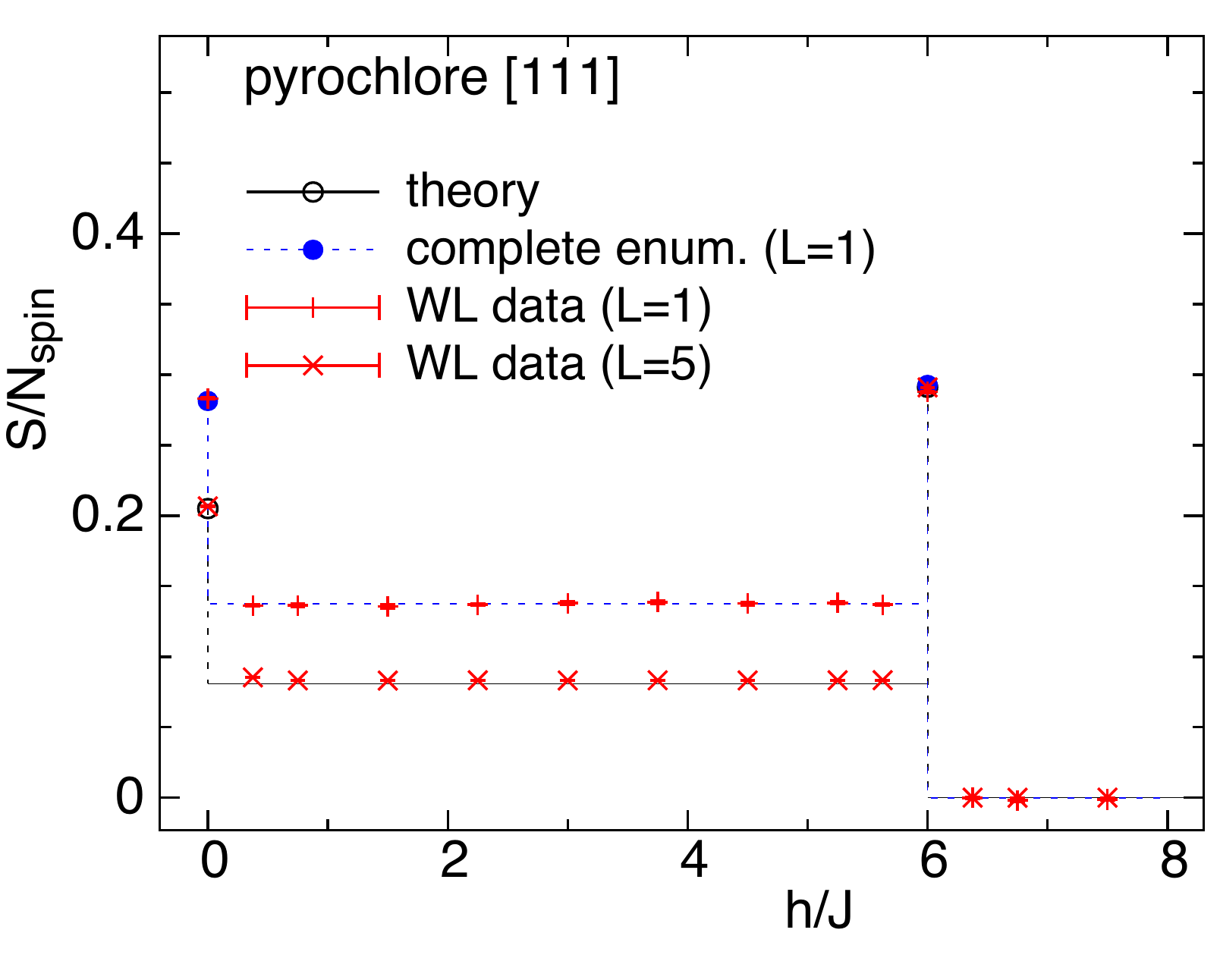}
\caption{
Comparison of the theory and the numerical results 
of the complete enumeration ($L=1$) and the WL method ($L=$~1~and~5) for the ground-state magnetizations and 
residual entropies per spin as a function of $h/J$ 
for the AFM Ising models on the pyrochlore lattice. 
The theoretical values are tabulated in Table II 
in the Appendix.
}
\label{fig:pyro_pure}
\end{center}
\end{figure*}
%%%%%%%%%%%%%%%%%%%%%%%%%%%%%%%%%%%%%%%%%%%%%%%%%%%%%%%%%%%%%%%%%%%%%%%%%%%%

In Fig.~\ref{fig:pyro_pure}, we plot the numerical results 
of the ground-state magnetization ($M/N$) and 
the residual entropy ($S/N$) of the AFM Ising model 
on the pyrochlore lattice for $L=$ 1 and 5. 
For WL data we took an average over 10 data. 
In the figure, we compare the numerical results 
with the theoretical values and complete enumeration. 
The ground state magnetization values obtained by the complete enumeration for $L = 1$ coinsides with the theory. 
There are two magnetization plateaus of $M/N=1/3$ and $M/N=1/2$. 
For $0<h/J<6$, which is the so-called "kagome-ice" state, the residual entropy 
is around 0.08, and there is no macroscopic degeneracy for $h/J>6$.  
At $h/J=6$, the two-in two-out configurations and 
the three-in one-out configurations coexist, and a large peak 
of the residual entropy is observed.  
The theoretical values are presented in Table II in the Appendix. 

%%%%%%%%%%%%%%%%%%%%%%%%%%%%%%%%%%%%%%%%%%%%%%%%%%%%%%%%%%%%%%%%%%%%%%%%%%%%
\begin{figure}
\begin{center}
\includegraphics[width=8.0cm]{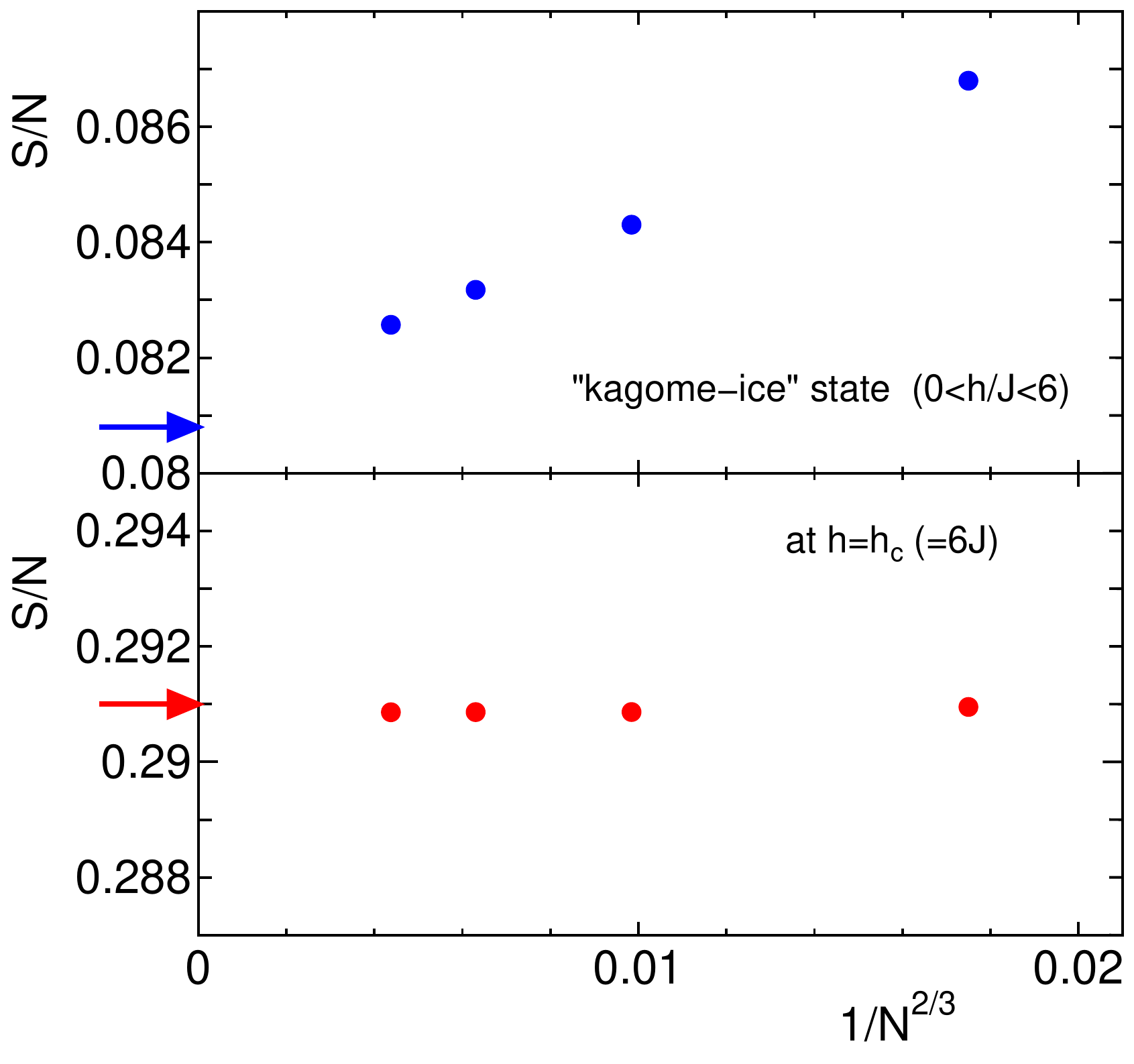}
\caption{
Plot of the residual entropy per spin 
of the pure AFM Ising model on the pyrochlore lattice 
under a magnetic field 
as a function of $1/N^{2/3}$ ($N=N_{\rm spin}$). 
The system size is $L=3 \ (N=432)$, 
$L=4 \ (N=1024)$, $L=5 \ (N=2000)$ and $L=6 \ (N=3456)$. 
The data for the "kagome-ice" state ($0<h/J<6$) are given 
by blue marks, whereas those at the crossover magnetic field 
($h/J=6$) are shown by red marks. 
The exact theoretical value for the "kagome-ice" 
state~\cite{Udagawa,Moessner} and the estimate for $h_c$ 
by the Bethe approximation~\cite{Nagle2,Isakov04} are shown 
by blue and red arrows, respectively. 
}
\label{fig:size_dep}
\end{center}
\end{figure}
%%%%%%%%%%%%%%%%%%%%%%%%%%%%%%%%%%%%%%%%%%%%%%%%%%%%%%%%%%%%%%%%%%%%%%%%%%%%

There is a small size dependence in the numerical estimate of the residual 
entropy for large enough systems. To check the accuracy of the calculation in detail, 
we examined the size dependence of the entropy in Fig.~\ref{fig:size_dep}. 
We utilized the data of $L=3$ ($N=432$), $L=4$ ($N=1024$), $L=5$ ($N=2000$), 
and $L=6$ ($N=3456$).  
We plot the residual entropy for the pure system 
as a function of $1/N^{2/3}$ ($N=N_{\rm spin}$) 
because this system is essentially a 2D one. 
We show the data of the "kagome-ice" state ($0<h/J<6$) and 
those of the crossover magnetic field ($h/J=6$). 
With respect to the data of the "kagome-ice" state, 
we averaged the values of $2*(3/8) \le h/J \le 14*(3/8)$.
The exact theoretical value for the "kagome-ice" 
state~\cite{Udagawa,Moessner} and the estimate for $h_c$ 
obtained by the Bethe approximation~\cite{Nagle2,Isakov04} are shown 
by blue and red arrows, respectively. 
We observe that our results agree well with the theoretical 
values of the residual entropy, $s_1/4=0.808$ for the "kagome-ice" 
state and $(3/4)s_5=0.291$ for $h_c$, which are 
presented in Table II in the Appendix.

In Figs.~\ref{fig:kagome_pure} and \ref{fig:tri_pure}, 
we also show the data of the pure models of the AFM Ising model 
on the kagome and triangular lattices in a magnetic field. 
We give the ground-state magnetizations per site and residual 
entropies per spin as a function of $h/J$. 
We obtained two magnetization plateaus and large entropy peaks 
between the two plateaus. 
The data obtained from the numerical simulation 
agree well with the theoretical values.

%%%%%%%%%%%%%%%%%%%%%%%%%%%%%%%%%%%%%%%%%%%%%%%%%%%%%%%%%%%%%%%%%%%%%%%%%%%%
\begin{figure*}
\begin{center}
\includegraphics[width=7.0cm]{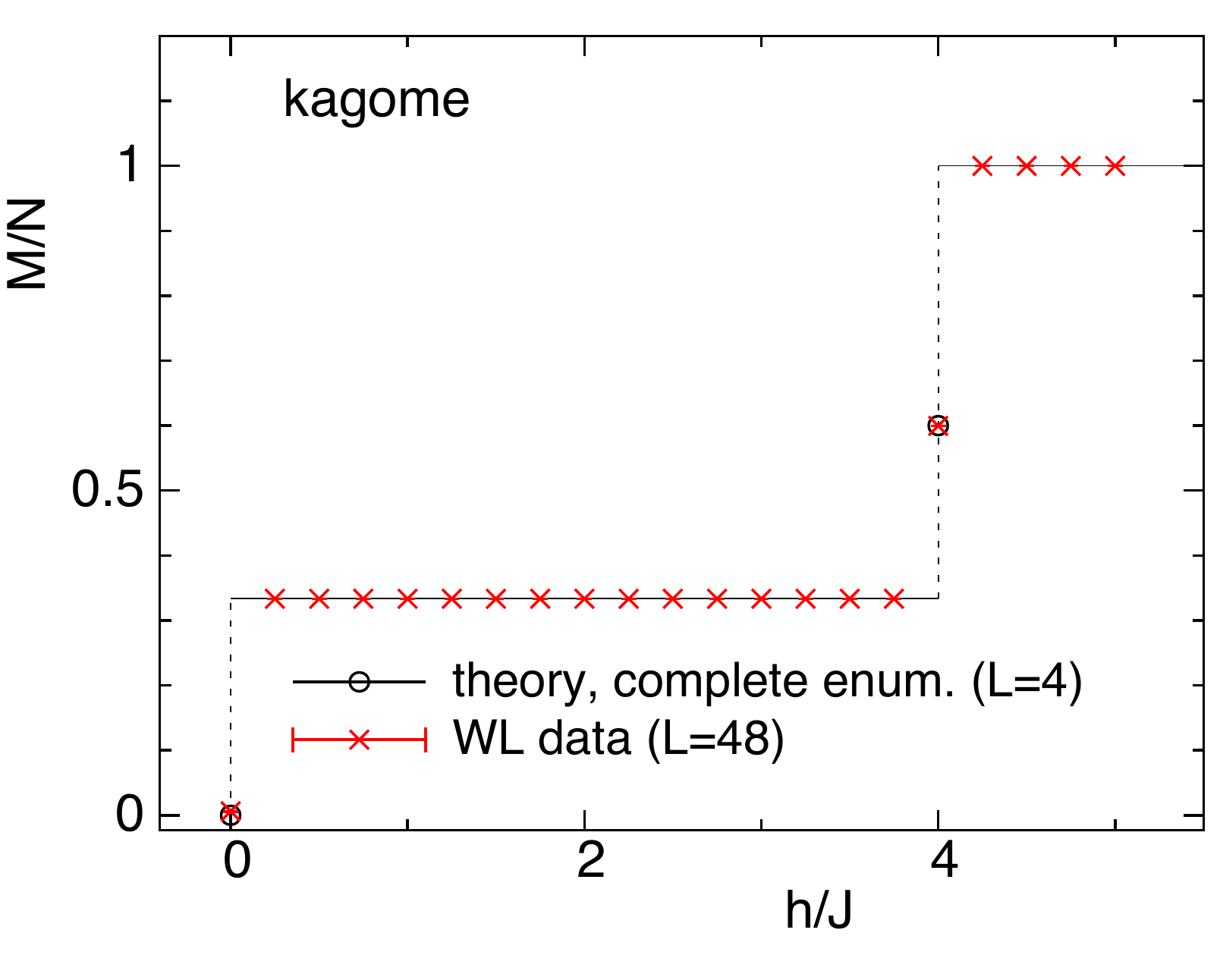}
\hspace{12mm}
\includegraphics[width=7.0cm]{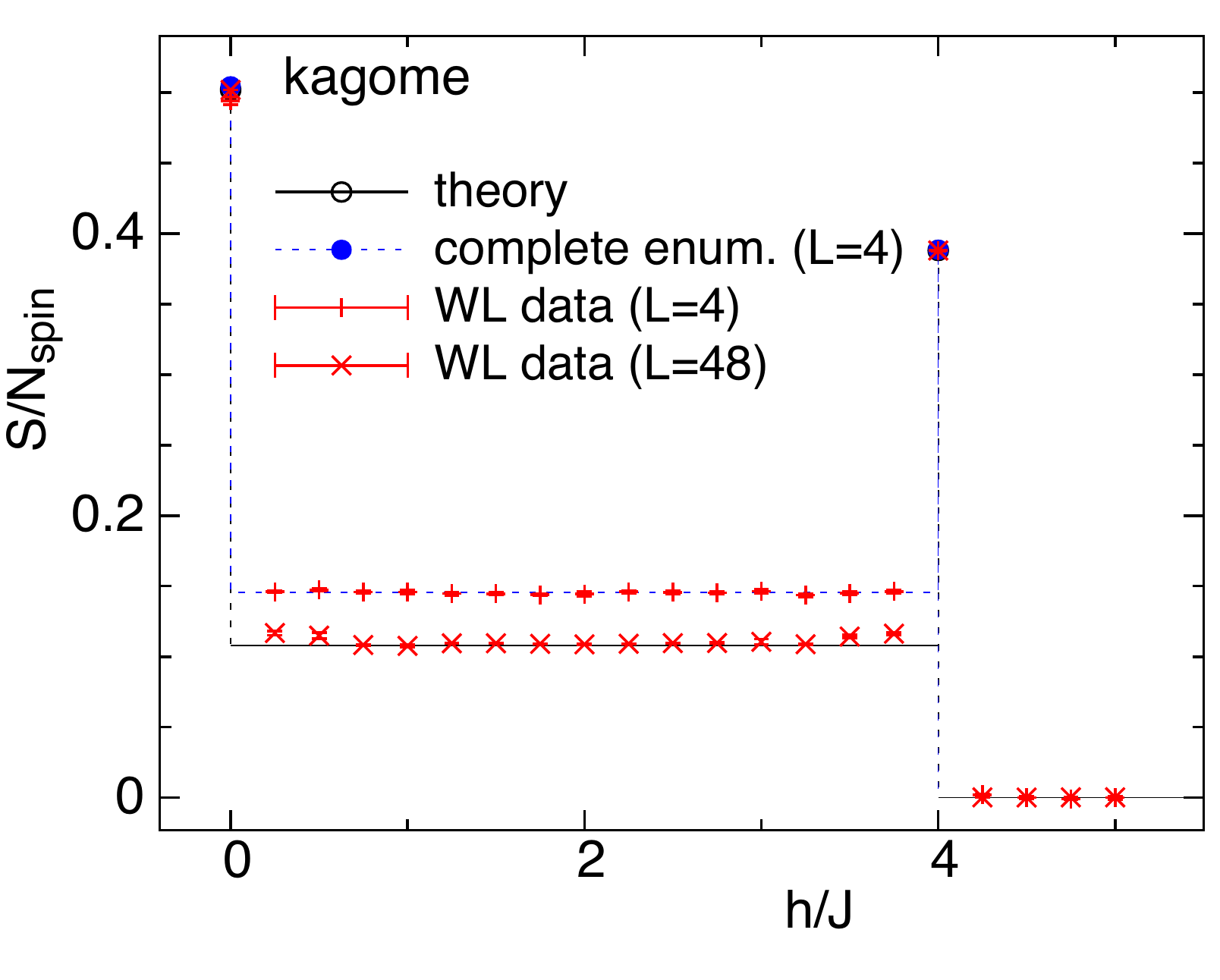}
\caption{
Comparison of the theory and the numerical results 
of the complete enumeration ($L=4$) and the WL method ($L=$~4~and~48)
for the ground-state magnetizations and residual 
entropies per spin as a function of $h/J$ 
for the AFM Ising models on the kagome lattice. 
The theoretical values are tabulated in Table II 
in the Appendix.
}
\label{fig:kagome_pure}
\end{center}
\end{figure*}
%%%%%%%%%%%%%%%%%%%%%%%%%%%%%%%%%%%%%%%%%%%%%%%%%%%%%%%%%%%%%%%%%%%%%%%%%%%%

%%%%%%%%%%%%%%%%%%%%%%%%%%%%%%%%%%%%%%%%%%%%%%%%%%%%%%%%%%%%%%%%%%%%%%%%%%%%
\begin{figure*}
\begin{center}
\includegraphics[width=7.0cm]{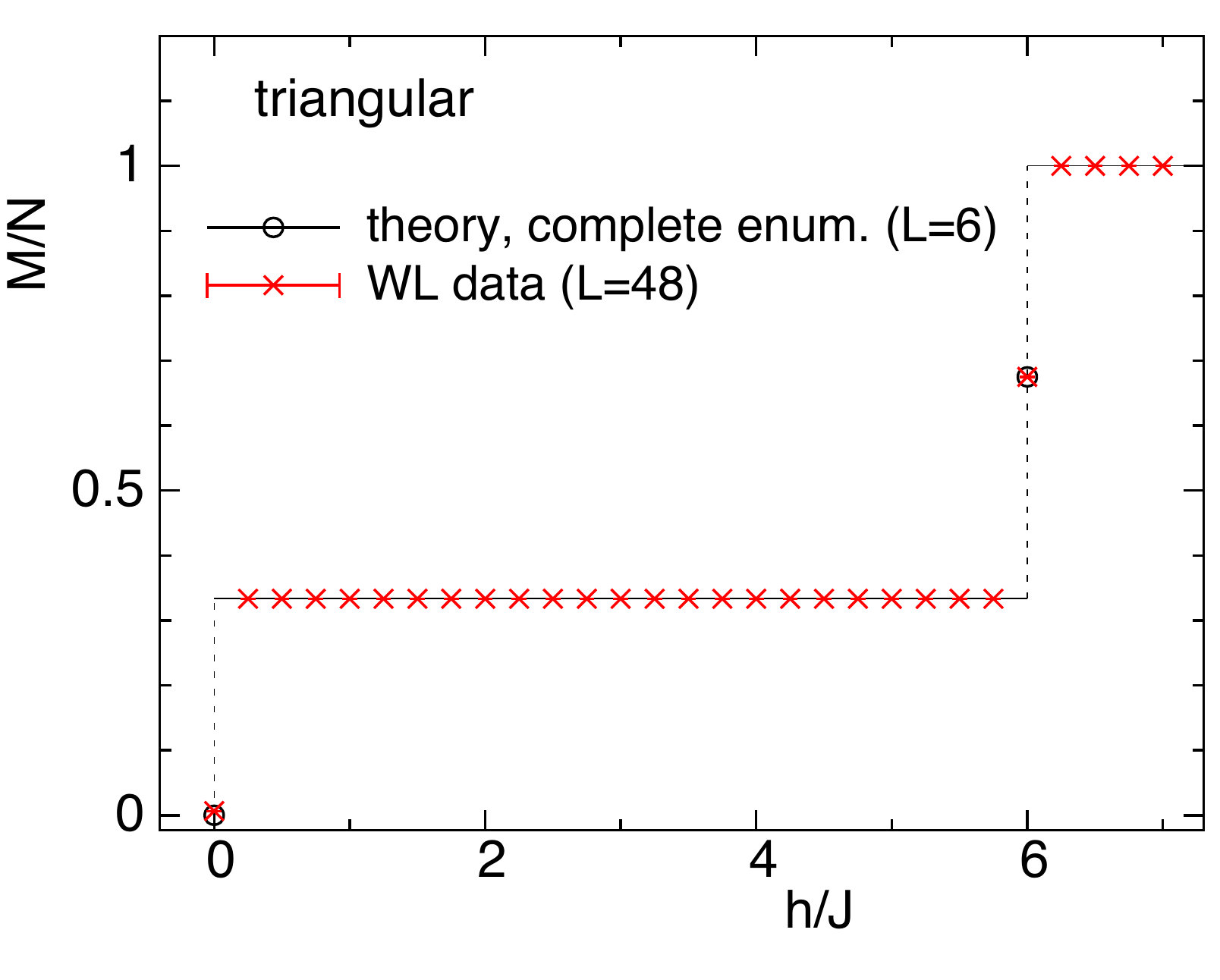}
\hspace{12mm}
\includegraphics[width=7.0cm]{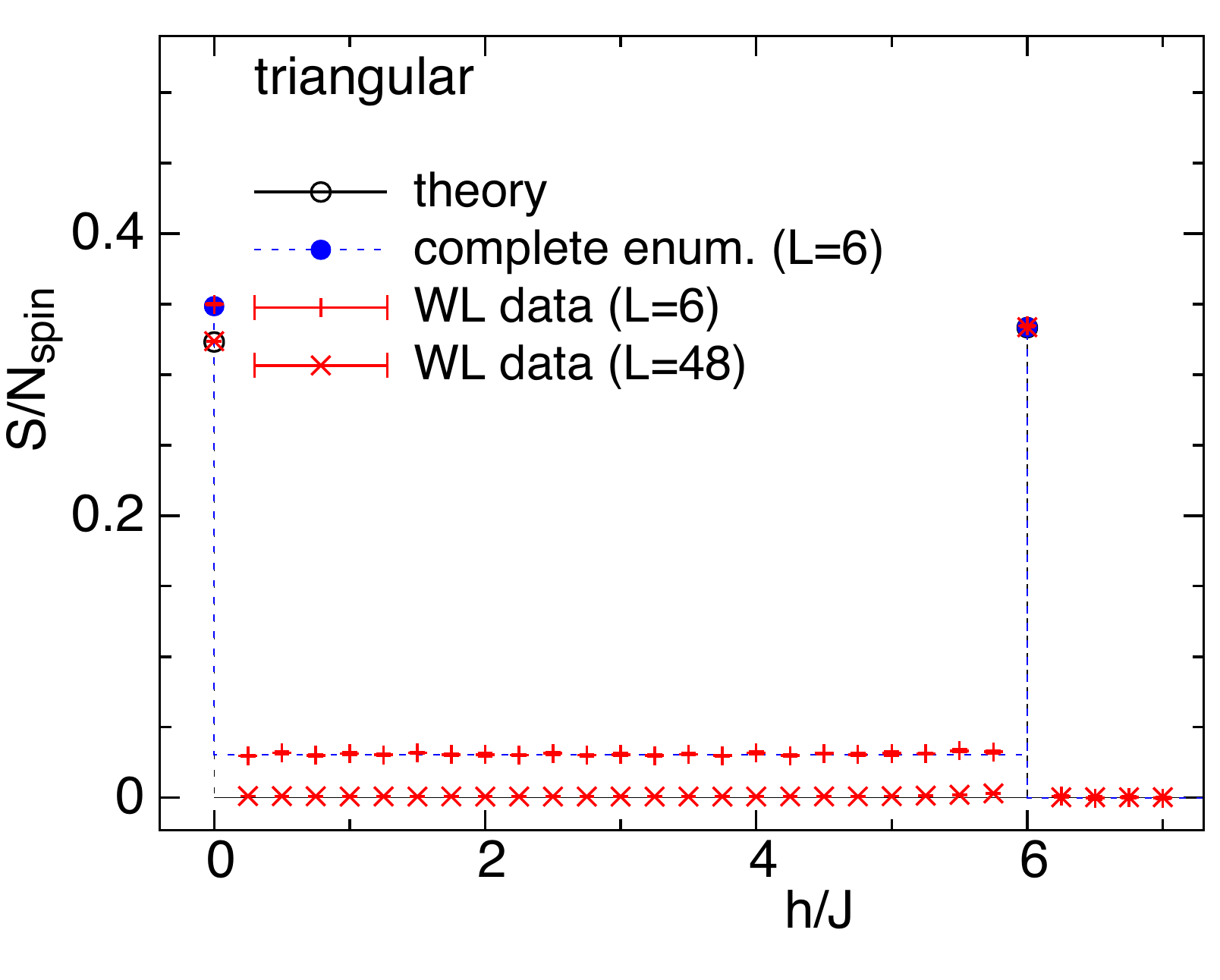}
\caption{
Comparison of the theory and the numerical results 
of the complete enumeration ($L=6$) and the WL method ($L=$~6~and~48)
of the WL method for the ground-state magnetizations and residual 
entropies per spin as a function of $h/J$ 
for the AFM Ising models on the triangular lattice. 
The theoretical values are tabulated in Table II 
in the Appendix.
}
\label{fig:tri_pure}
\end{center}
\end{figure*}
%%%%%%%%%%%%%%%%%%%%%%%%%%%%%%%%%%%%%%%%%%%%%%%%%%%%%%%%%%%%%%%%%%%%%%%%%%%%

\subsection{Diluted model on the pyrochlore lattice}

\begin{figure*}
\begin{center}
\includegraphics[width=7.0cm]{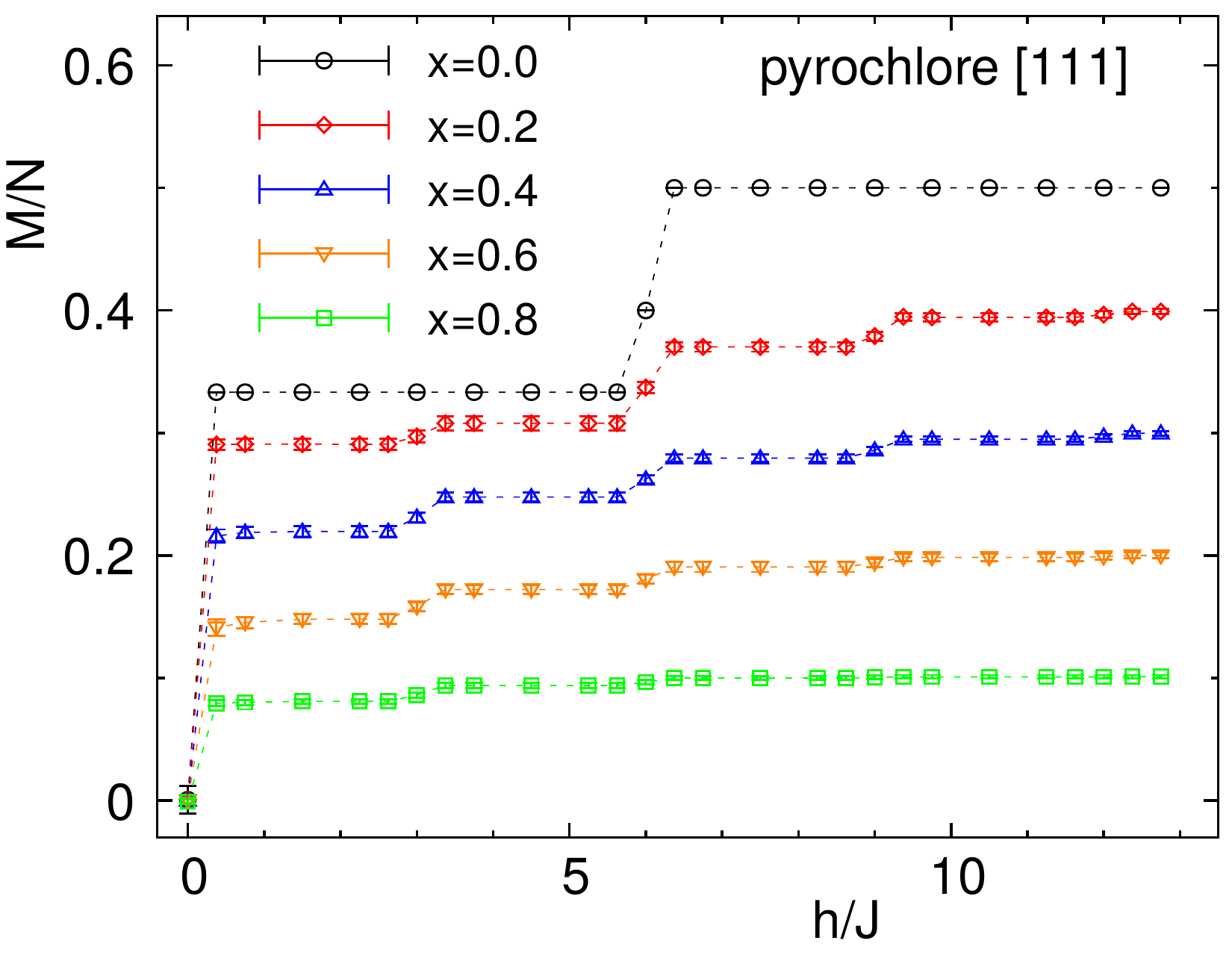}
\hspace{12mm}
\includegraphics[width=7.0cm]{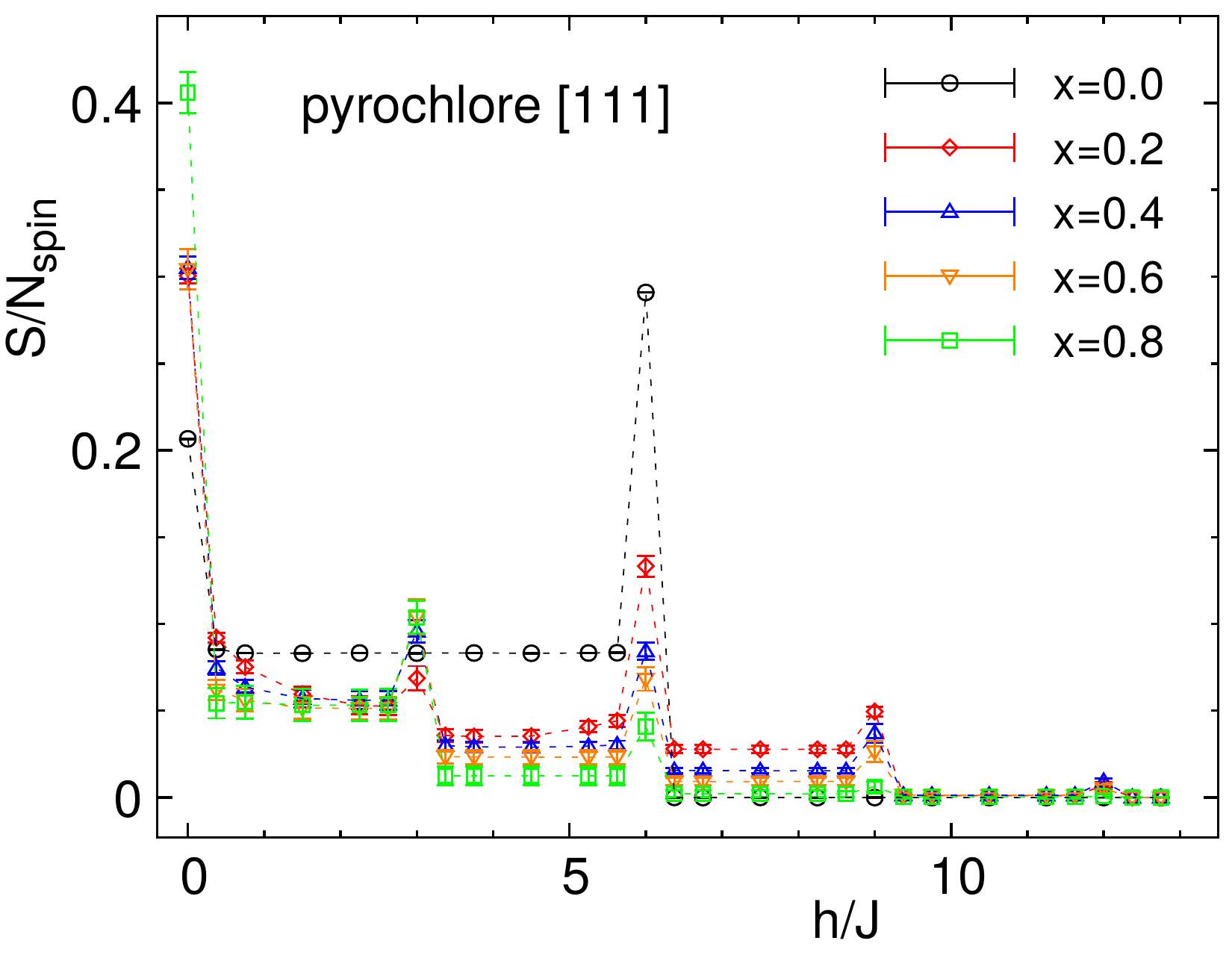}
\caption{
Magnetic-field ($h$) dependence of the ground-state magnetization 
per site (left) and the residual entropy per spin (right)
of the AFM Ising model on the pyrochlore lattice. 
The magnetic field is applied in the [111] direction. 
The system size is $L=5$ ($N=2000$). 
The dilution concentrations are $x$ = 0.0, 0.2, 0.4, 0.6, and 0.8.
}
\label{fig:pyro_dilute}
\end{center}
\end{figure*}

Next, we focus on the dilution effect, which is the main subject of this paper. 
In Fig.~\ref{fig:pyro_dilute}, we plot the magnetic-field ($h$) dependence 
of the ground-state magnetization per site and the residual entropy per spin 
of the diluted AFM Ising model on the pyrochlore lattice. 
The system size is $L=5$ ($N=2000$). 
The average was taken over 10 random samples.
With the scale of this plot, the size dependence is observed to be small. 
The dilution concentrations are $x$ = 0.0, 0.2, 0.4, 0.6, and 0.8. 

We observe a stepwise increase of the magnetization for the diluted case 
($x \ne 0$).  For the diluted case, there are five magnetization plateaus 
instead of two, which is consistent with the result of the previous 
replica-exchange Monte Carlo study~\cite{Peretyatko}. 
The magnetization steps are observed at $h/J$ = 3, 6, 9, and 12, 
which are contrary to the pure case in that there is 
only a magnetization step at $h/J=6$. 
The saturated values of the magnetization per site become $(1/2)*(1-x)$.

In the right figure of Fig.~\ref{fig:pyro_dilute}, we see a stepwise decrease 
of the residual entropy as a function of the magnetic field $h$. 
The results for the case with no magnetic field ($h=0$) are of course 
the same as the previous study~\cite{Shevchenko}. 
The value for $0<h/J<3$ and that for $3<h/J<6$ 
are different, and a peak appears at $h/J=3$, 
where there is no peak for a pure system ($x=0$). 
We observe another large peak at $h/J=6$, as in a pure system, 
and the residual entropy for $6<h/J<9$ is not zero. 
There is a peak at $h/J=9$, and the residual entropy 
becomes zero for $h/J>9$.  We observe a small peak at $h/J=12$. 
The positions of the peaks in entropy, at $h/J$ = 3, 6, 9, 
and 12, are where the magnetization steps appear.  
At the crossover magnetic field $h_c$, the states 
with different magnetization are degenerate,
which results in the large peak of the residual entropy. 
It is interesting to note that there are nonzero residual entropies 
at $6<h/J<9$ for the diluted case, although the residual entropy 
is zero for the pure case. We may call this phenomenon as 
the residual entropy induced by dilution. 

There is a dilution concentration ($x$) dependence in the behavior 
of the ground-state magnetization and the residual entropy. 
We can understand this dependence from an analysis of the origin 
of the five magnetization plateaus given in the previous 
study~\cite{Peretyatko}.  The energy analysis at the spin flip 
for two corner-sharing tetrahedra is given in Fig.~6 
in~\cite{Peretyatko}, and the proportions of the types of 
spin configuration in the tetrahedron were investigated, 
as shown in Fig.~5 in~\cite{Peretyatko}. 

To summarize this subsection, we obtain the same multiple 
magnetization plateaus as the previous replica-exchange Monte Carlo 
study~\cite{Peretyatko}. We observe peaks of the entropy 
at $h/J$ = 3, 6, 9, and 12, which correspond to the positions of 
the magnetization steps. 
The large peak at $h/J=6$ is the same as the pure case, 
but other peaks appear only for diluted models. 
The peak at $h/J=12$ is very weak. 
We sometimes observe the residual entropy induced by dilution.

%%%%%%%%%%%%%%%%%%%%%%%%%%%%%%%%%%%%%%%%%%%%%%%%%%%%%%%%%%%%%%%%%%%%%%%%%%%%
%%%%%%%%%%%%%%%%%%%%%%%%%%%%%%%%%%%%%%%%%%%%%%%%%%%%%%%%%%%%%%%%%%%%%%%%%%%%

\subsection{Diluted model on the kagome lattice}
%%%%%%%%%%%%%%%%%%%%%%%%%%%%%%%%%%%%%%%%%%%%%%%%%%%%%%%%%%%%%%%%%%%%%%%%%%%%
\begin{figure*}
\begin{center}
\includegraphics[width=7.0cm]{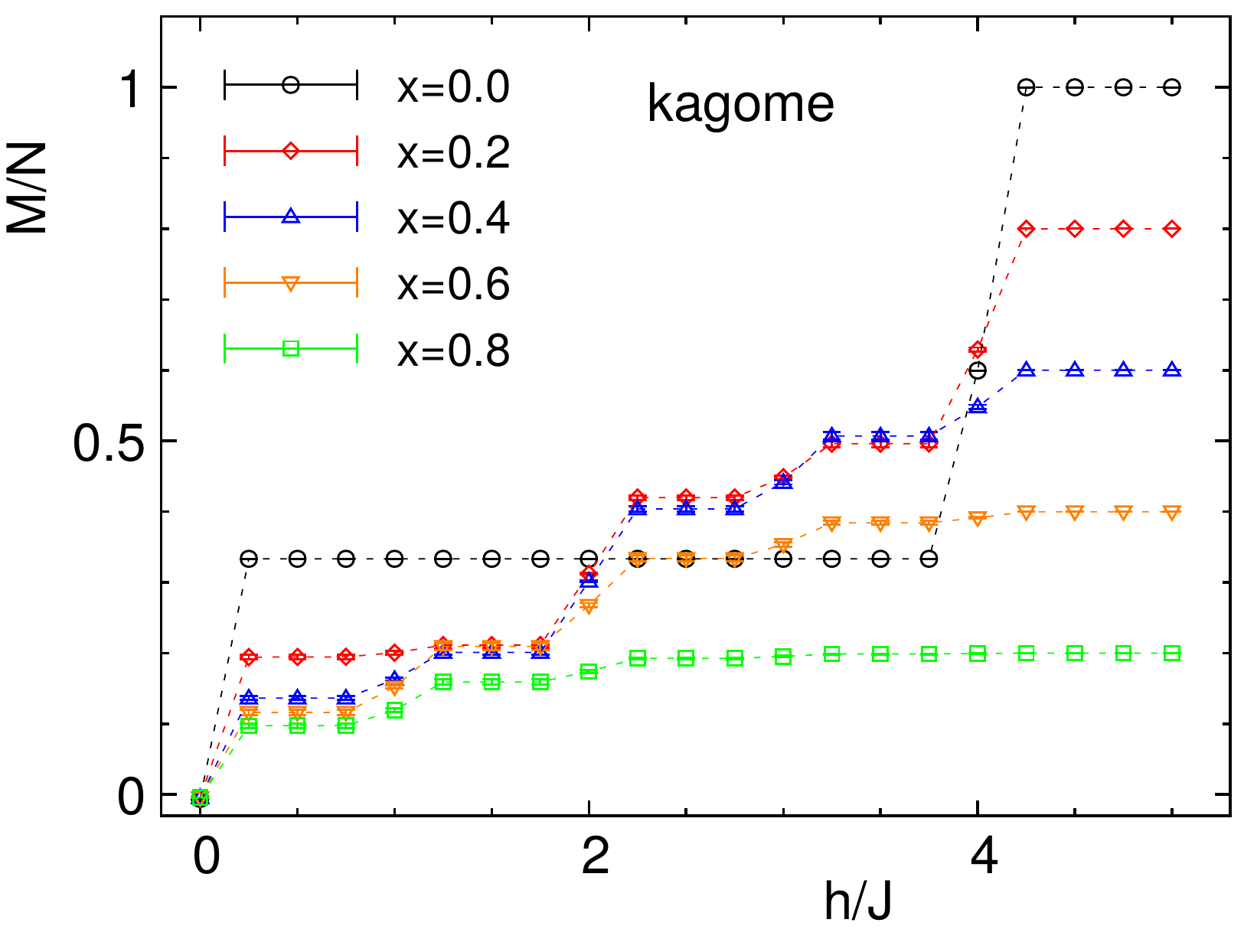}
\hspace{12mm}
\includegraphics[width=7.0cm]{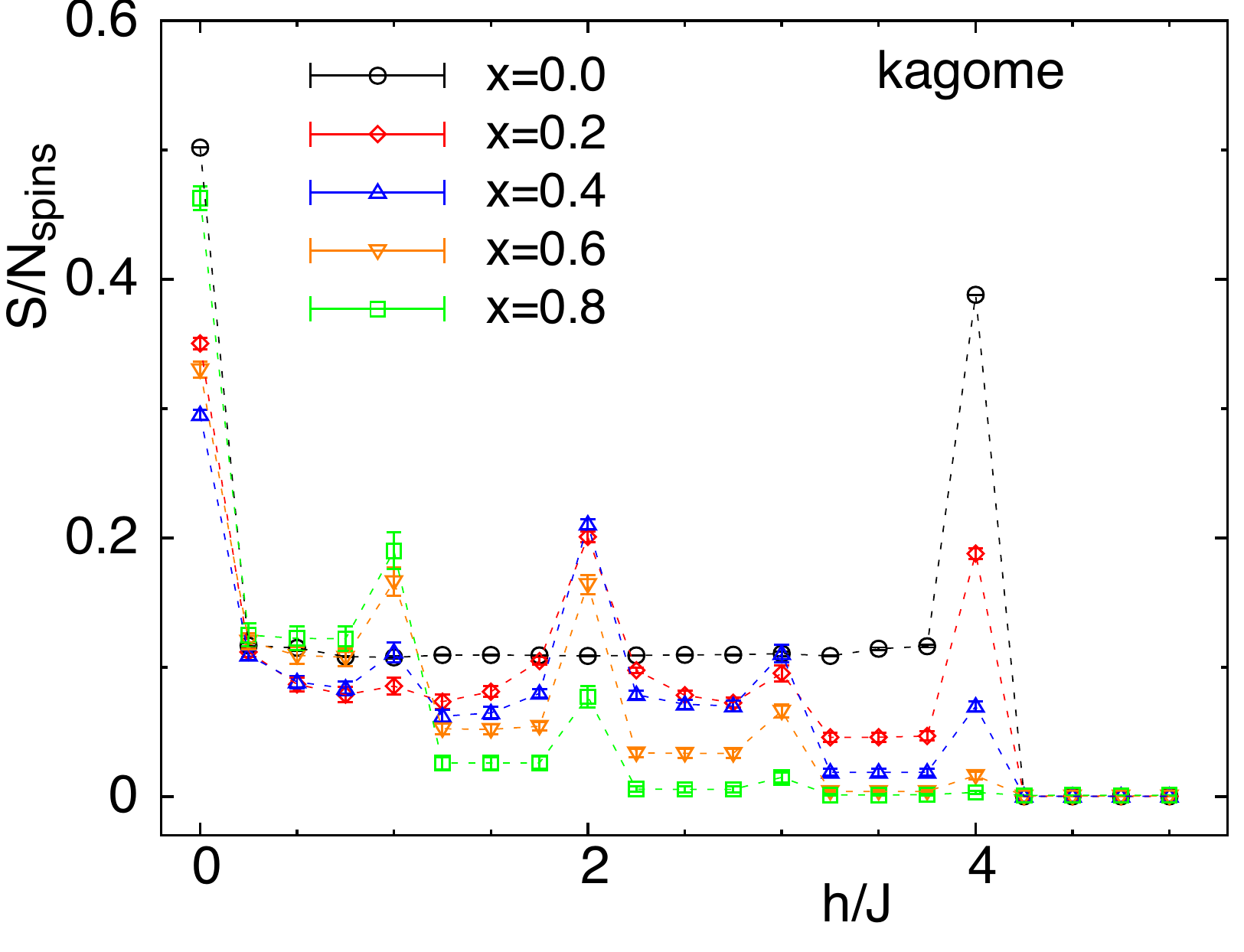}
\caption{
Magnetic-field ($h$) dependence of the ground-state magnetization 
per site (left) and the residual entropy per spin (right)
of the AFM Ising model on the kagome lattice. 
The system size is $L=48$ ($N=3456$). 
The dilution concentrations are $x$ = 0.0, 0.2, 0.4, 0.6, and 0.8.
}
\label{fig:kagome_dilute}
\end{center}
\end{figure*}
%%%%%%%%%%%%%%%%%%%%%%%%%%%%%%%%%%%%%%%%%%%%%%%%%%%%%%%%%%%%%%%%%%%%%%%%%%%%

In Fig.~\ref{fig:kagome_dilute}, we plot the magnetic-field ($h$) dependence 
of the magnetization per site and the residual entropy per spin 
of the diluted AFM Ising model on the kagome lattice. 
The system size is $L=48$ ($N=3456$). 
The dilution concentrations are $x$ = 0.0, 0.2, 0.4, 0.6, and 0.8.

We observe five plateaus in the magnetization, 
which is the same as the pyrochlore lattice. 
The saturated magnetization is $N*(1-x)$. 
There are magnetization steps at $h/J$ = 1, 2, and 3 in addition 
to $h/J=4$ of the pure case.  The magnetic fields of additional steps 
are smaller than that of the pure model, which is different 
from the situation of the pyrochlore lattice. 
The magnetization plateaus result from the competition 
between the exchange and Zeeman energies. 
We can understand the origin of the multiple magnetization 
plateaus using the same analysis as in the case of 
the pyrochlore lattice~\cite{Peretyatko}. 
We consider the spin configuration in the triangle 
for the kagome lattice instead of the tetrahedron. 
The detailed analysis of the similarities and dissimilarities 
between the magnetization curves of the diluted model 
for the "kagome-ice" state and for the kagome lattice given in~\cite{Soldatov} 
together with the replica-exchange Monte Carlo simulation 
of the model on the kagome lattice.

We observe peaks of the residual entropy at $h/J$ 
= 1, 2, 3, and 4, where the magnetization steps appear. 
The states with different magnetization are degenerate at the crossover magnetic fields $h_c$.  
The origin of the large peaks in the entropy is the same as 
the case of the pyrochlore lattice. 

\subsection{Diluted model on the triangular lattice}

%%%%%%%%%%%%%%%%%%%%%%%%%%%%%%%%%%%%%%%%%%%%%%%%%%%%%%%%%%%%%%%%%%%%%%%%%%%%
\begin{figure*}
\begin{center}
\includegraphics[width=7.0cm]{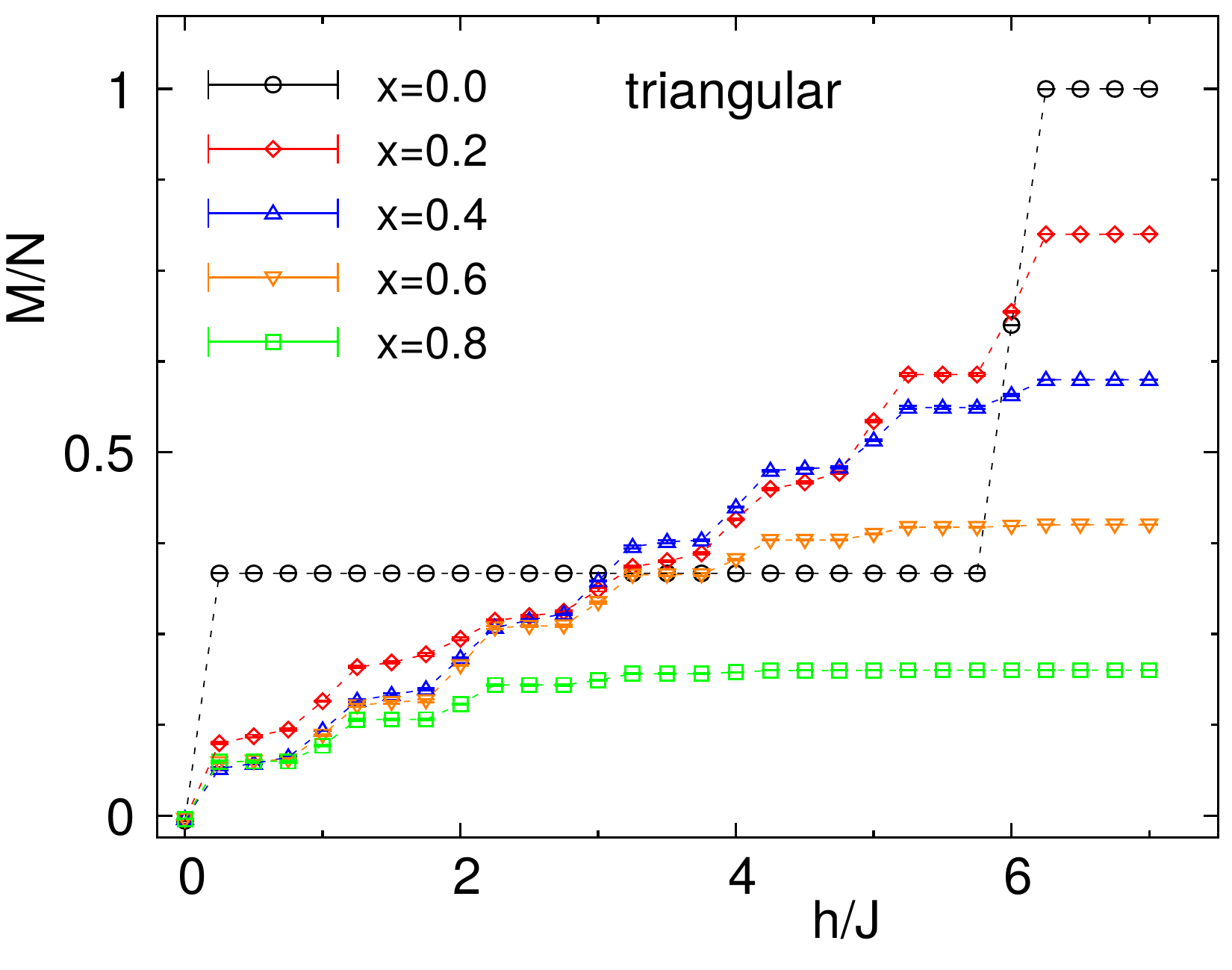}
\hspace{12mm}
\includegraphics[width=7.0cm]{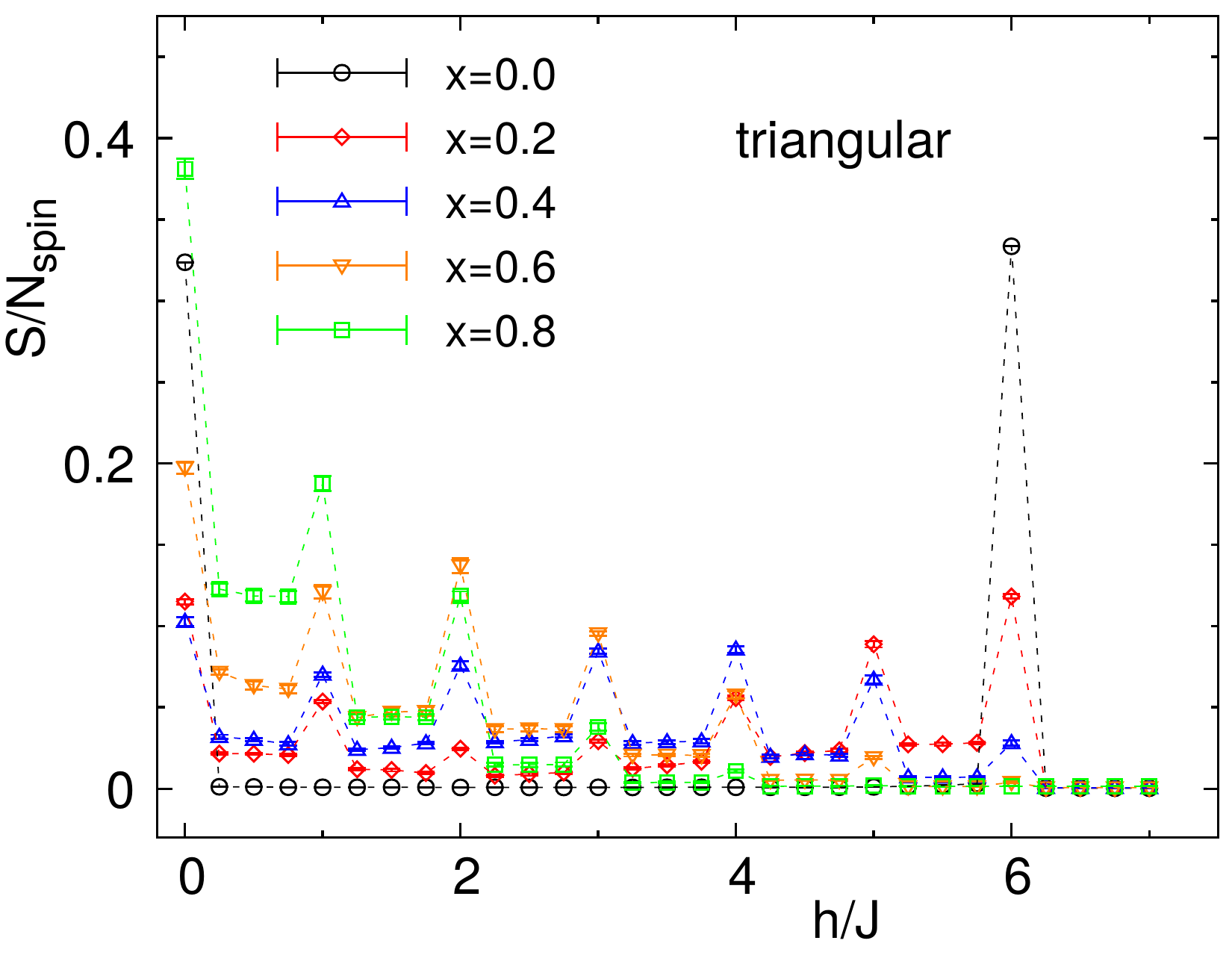}
\caption{
Magnetic-field ($h$) dependence of the ground-state magnetization 
per site (left) and the residual entropy per spin (right)
of the AFM Ising model on the triangular lattice. 
The system size is $L=48$ ($N=2304$). 
The dilution concentrations are $x$ = 0.0, 0.2, 0.4, 0.6, and 0.8.
}
\label{fig:tri_dilute}
\end{center}
\end{figure*}
%%%%%%%%%%%%%%%%%%%%%%%%%%%%%%%%%%%%%%%%%%%%%%%%%%%%%%%%%%%%%%%%%%%%%%%%%%%%

In Fig.~\ref{fig:tri_dilute}, we present a plot of the magnetic-field ($h$) 
dependence of the magnetization per site and the residual entropy per spin 
of the diluted AFM Ising model on the triangular lattice. 
The system size is $L=48$ ($N=2304$). 
The dilution concentrations are $x$ = 0.0, 0.2, 0.4, 0.6, and 0.8.

Seven magnetization plateaus are observed. 
New magnetization steps appear at $h/J$ = 1, 2, 3, 4, 
and 5 in addition to $h/J=6$ of the pure case.  
The saturated magnetization is $N*(1-x)$. 
Such multiple magnetization plateaus were previously reported 
for weak magnetic fields~\cite{Yao,Zukovic}. 
We can understand the origin of the multiple 
magnetization plateaus in terms of the spin configuration 
in the basic unit of triangle, which can be found in~\cite{Soldatov}.

We find a nonzero entropy for the diluted model for $h/J<6$, 
which can be regarded as the residual entropy due to the dilution.  
We again observe peaks of the residual entropy at $h/J$ 
= 1, 2, 3, 4, 5 and 6, which correspond to the positions at which 
the magnetization steps appear. 
The states with different magnetization are degenerate at the crossover magnetic fields $h_c$,
 which is a common origin of the large peaks of the entropy. 

\subsection{Explanation of the origin of large entropy peaks and magnetizations plateaus in the pyrochlore lattice}
\label{origin_section}
  %%%%%%%%%%%%%%%%%%%%%%%%%%%%%%%%%%%%%%%%%%%%%%%%%%%%%%%%%%%%%%%%%%%%%%%%%%%%

\begin{figure*}
	\begin{minipage}[h]{0.3\linewidth}
		\center{\includegraphics[width=1\linewidth]{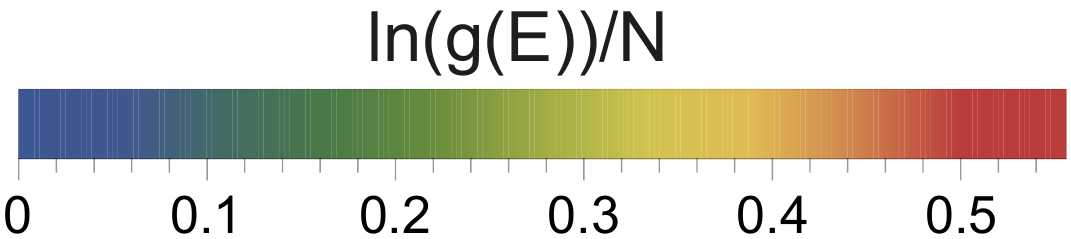}} 
	\end{minipage}
	\vfill
	\begin{minipage}[h]{0.4\linewidth}
		\center{\includegraphics[width=1\linewidth]{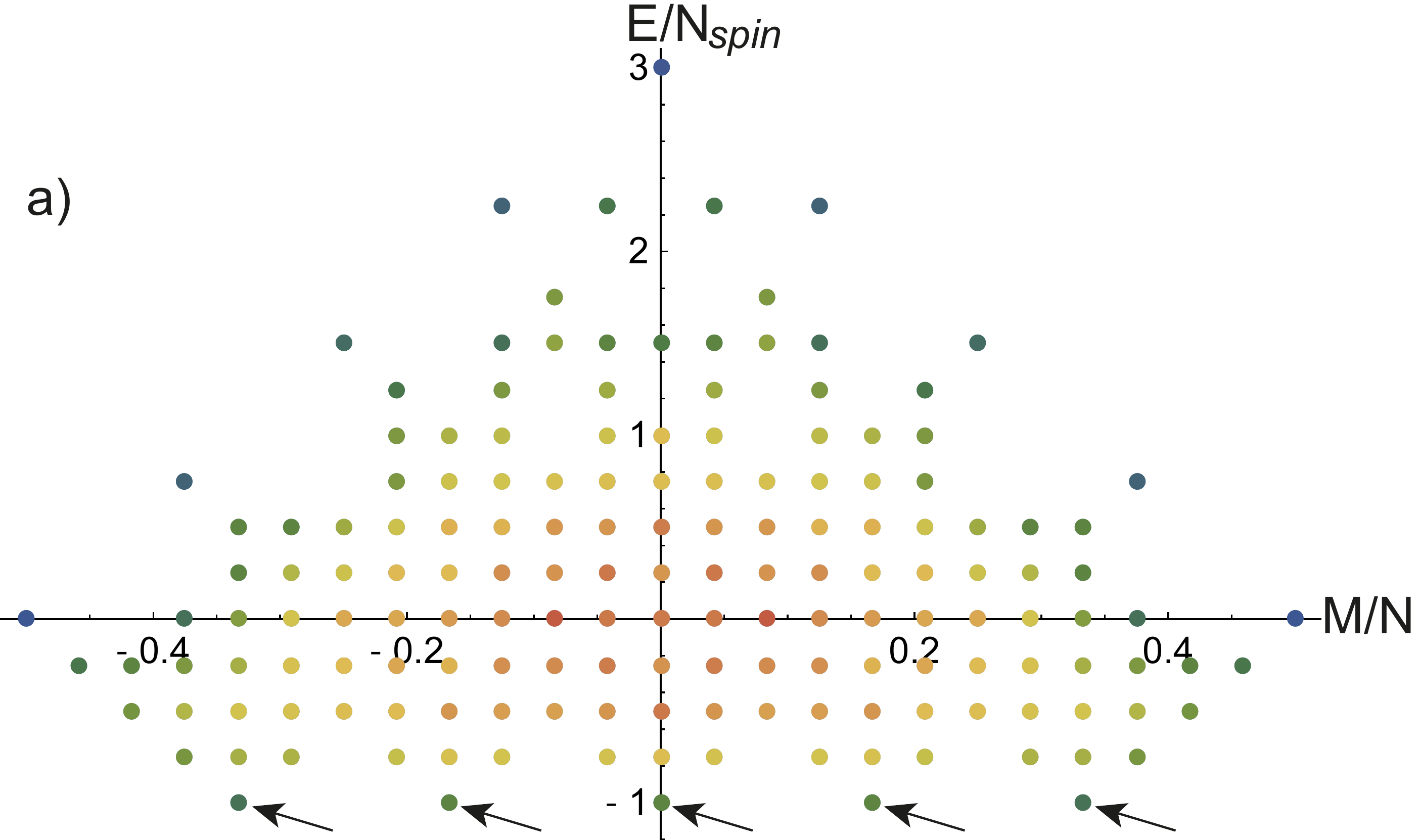}}
	\end{minipage}
	\hfill $h/J=0$ \hfill 
	\begin{minipage}[h]{0.4\linewidth}
		\center{\includegraphics[width=1\linewidth]{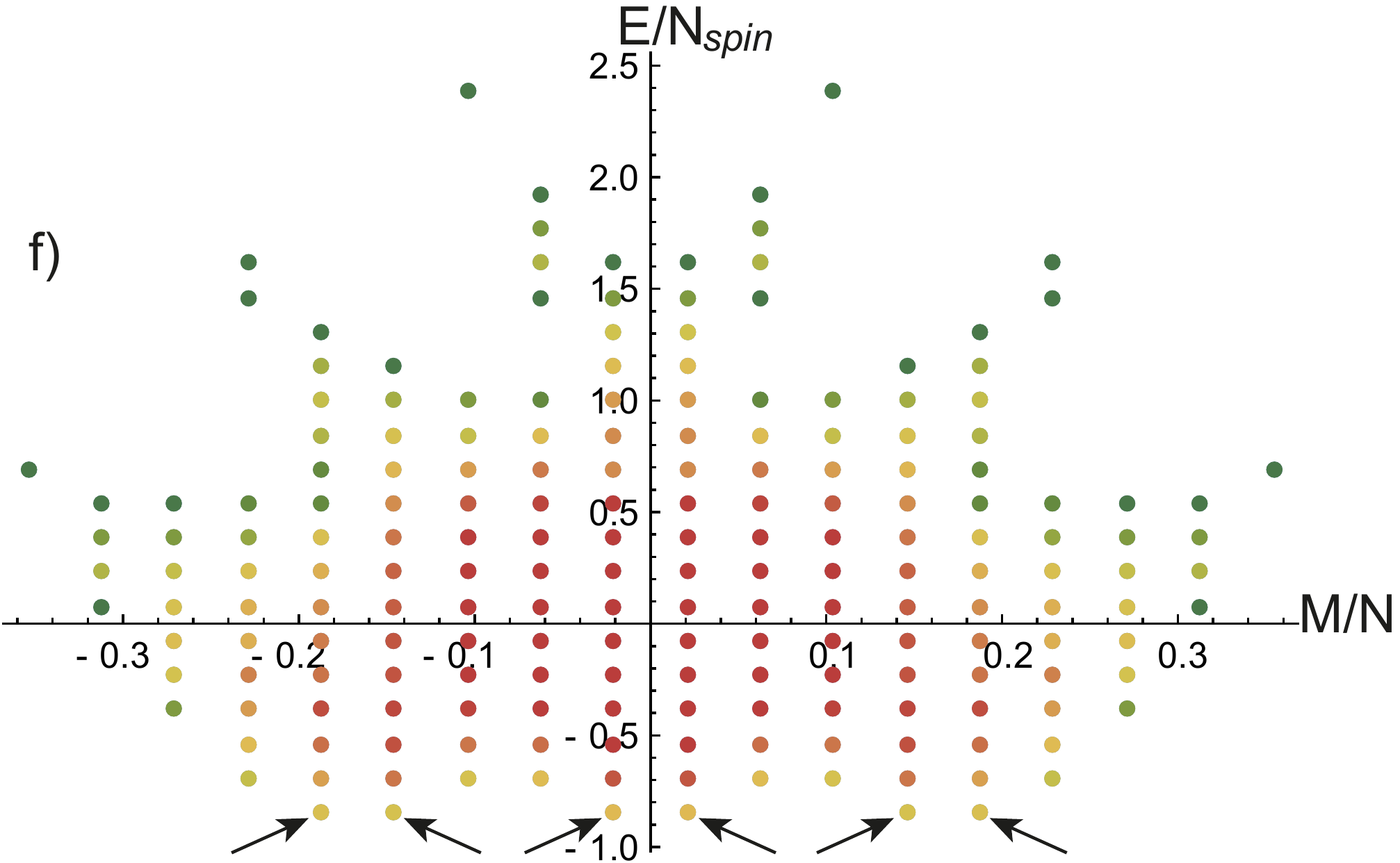}}
	\end{minipage}
	\vfill
	\begin{minipage}[h]{0.4\linewidth}
	\center{\includegraphics[width=1\linewidth]{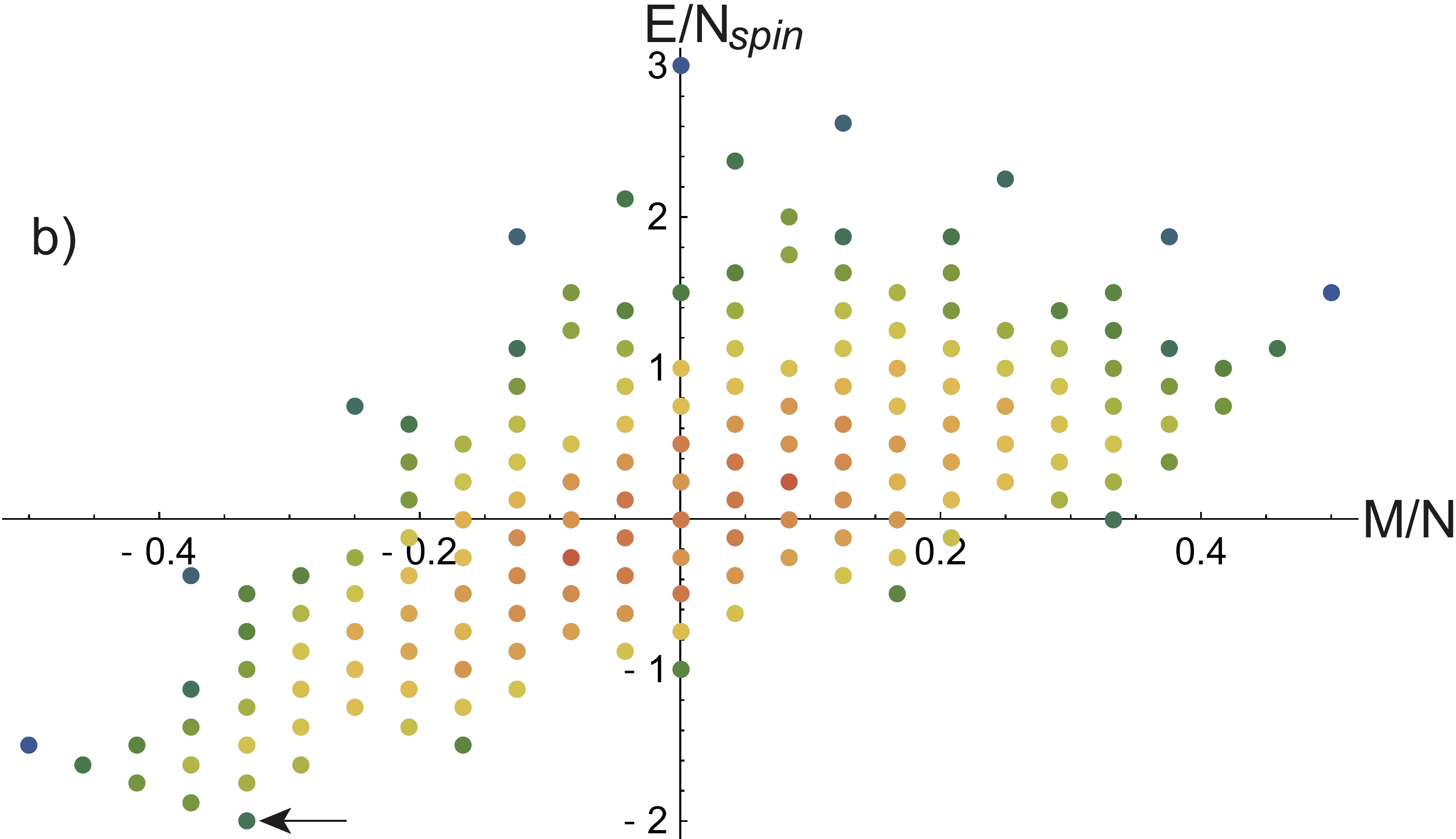}}
	\end{minipage}
	\hfill $h/J=3$ \hfill
	\begin{minipage}[h]{0.4\linewidth}
	\center{\includegraphics[width=1\linewidth]{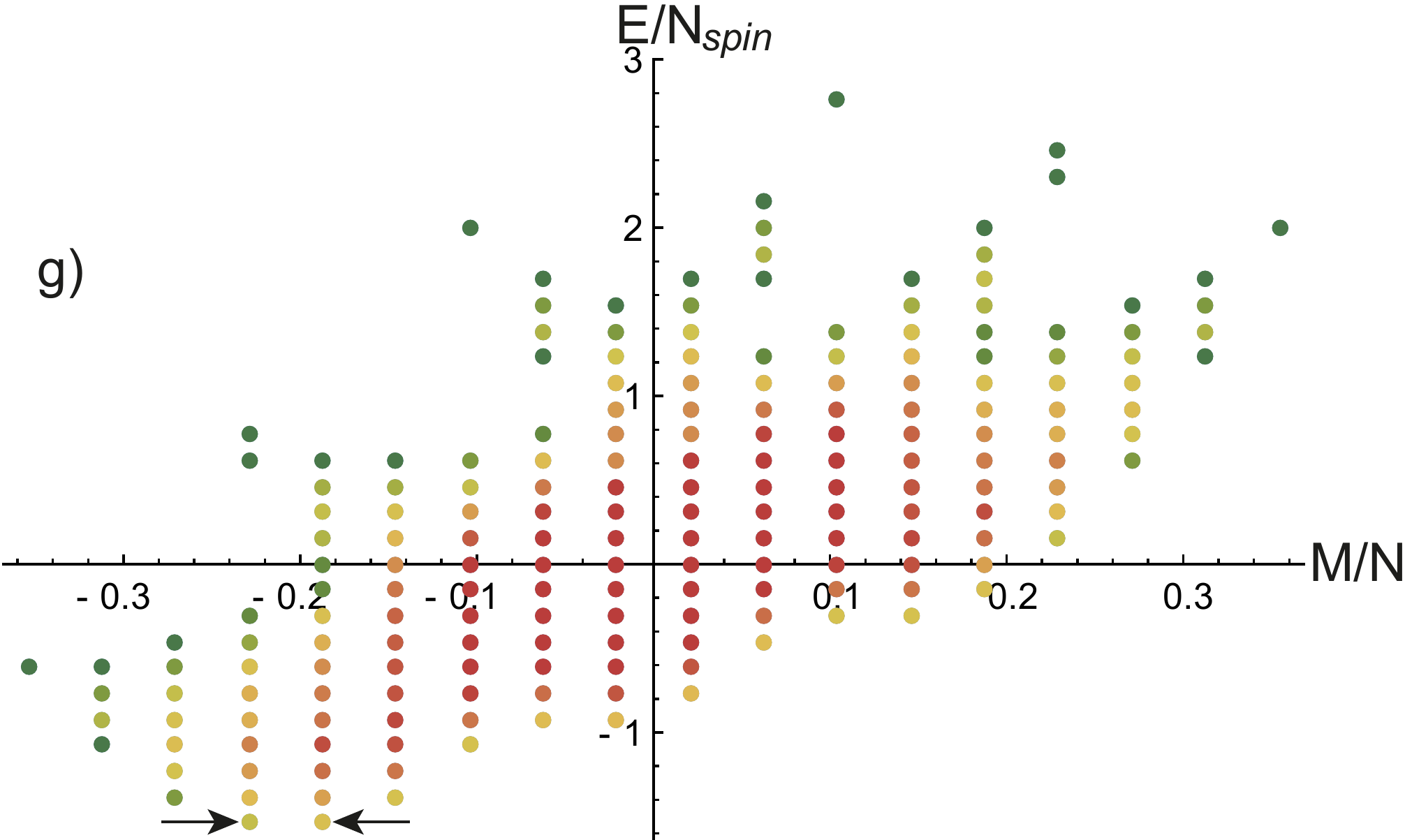}}
	\end{minipage}
	\vfill
	\begin{minipage}[h]{0.4\linewidth}
	\center{\includegraphics[width=1\linewidth]{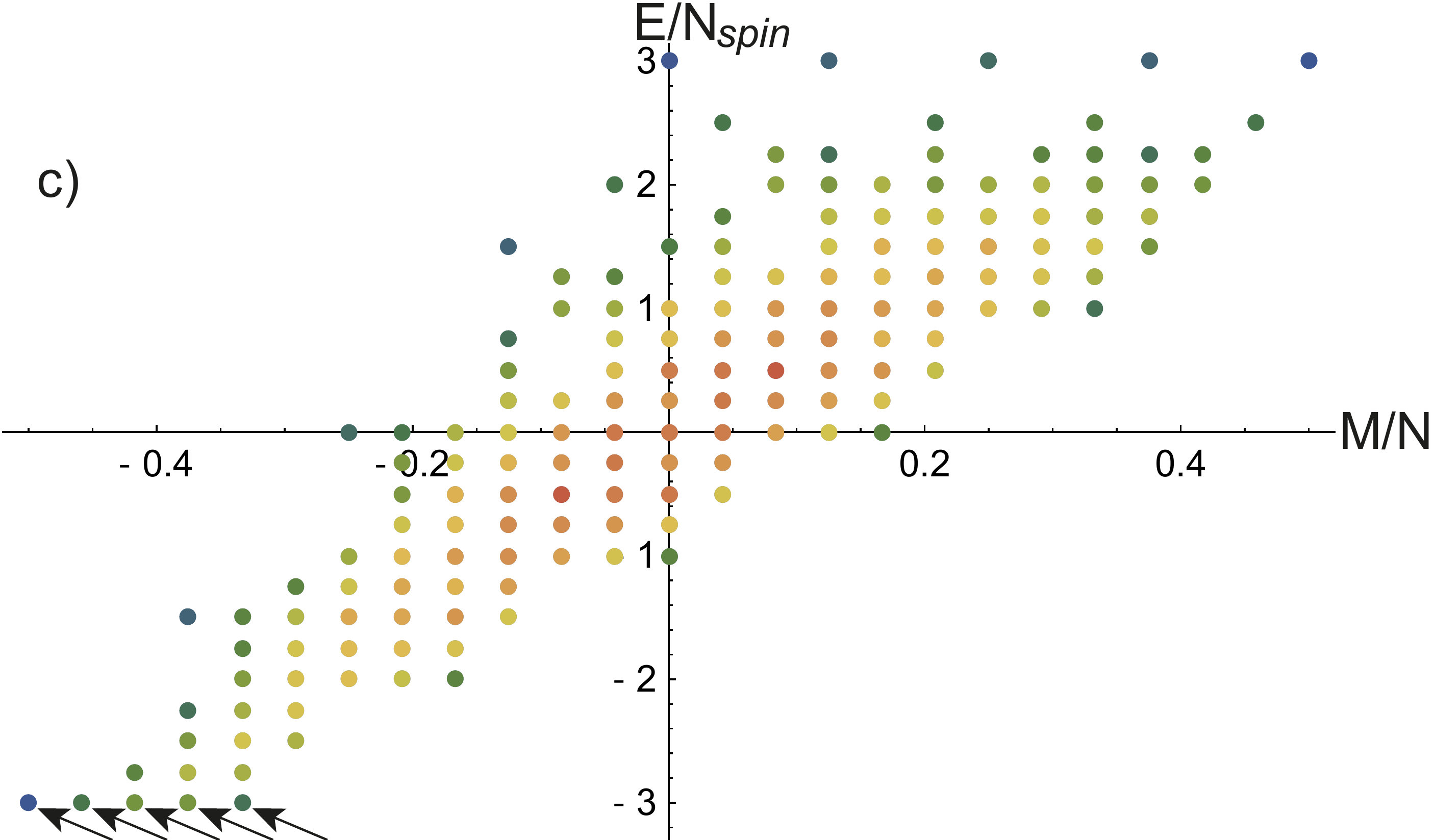}} 
	\end{minipage}
	\hfill $h/J=6$ \hfill
	\begin{minipage}[h]{0.4\linewidth}
	\center{\includegraphics[width=1\linewidth]{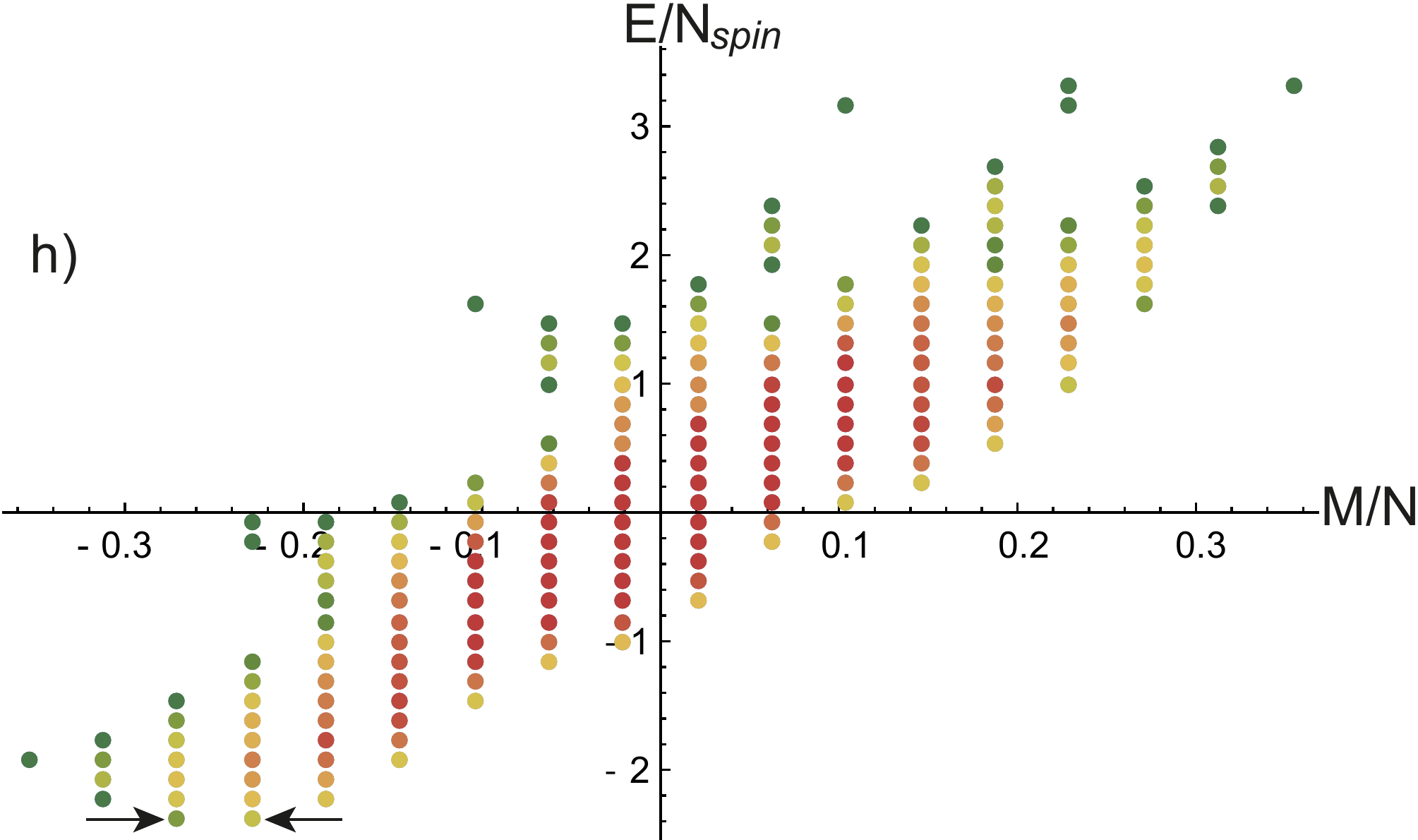}}
	\end{minipage}
	\vfill
	\begin{minipage}[h]{0.4\linewidth}
	\center{\includegraphics[width=1\linewidth]{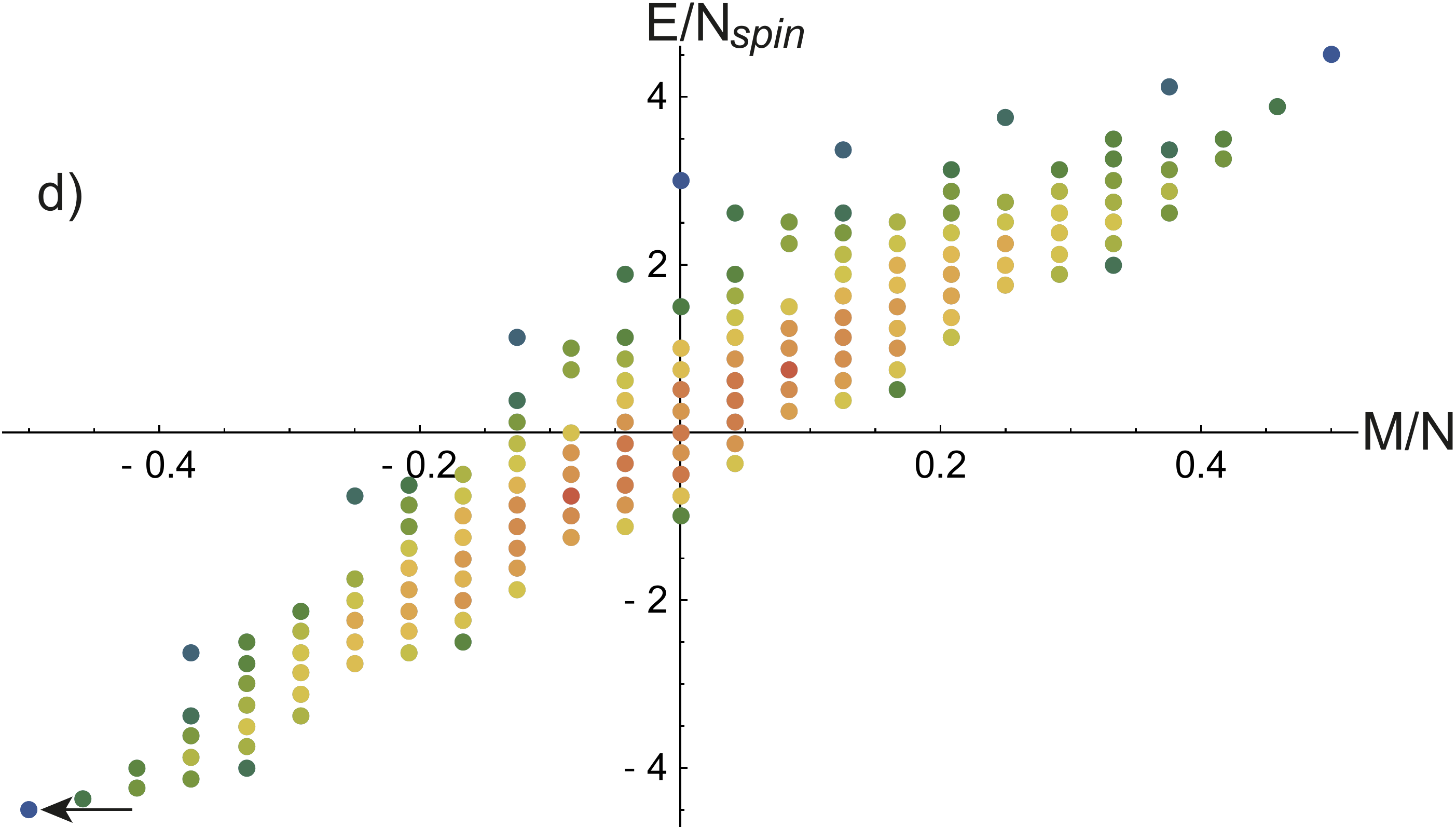}} 
	\end{minipage}
	\hfill $h/J=9$ \hfill
	\begin{minipage}[h]{0.4\linewidth}
	\center{\includegraphics[width=1\linewidth]{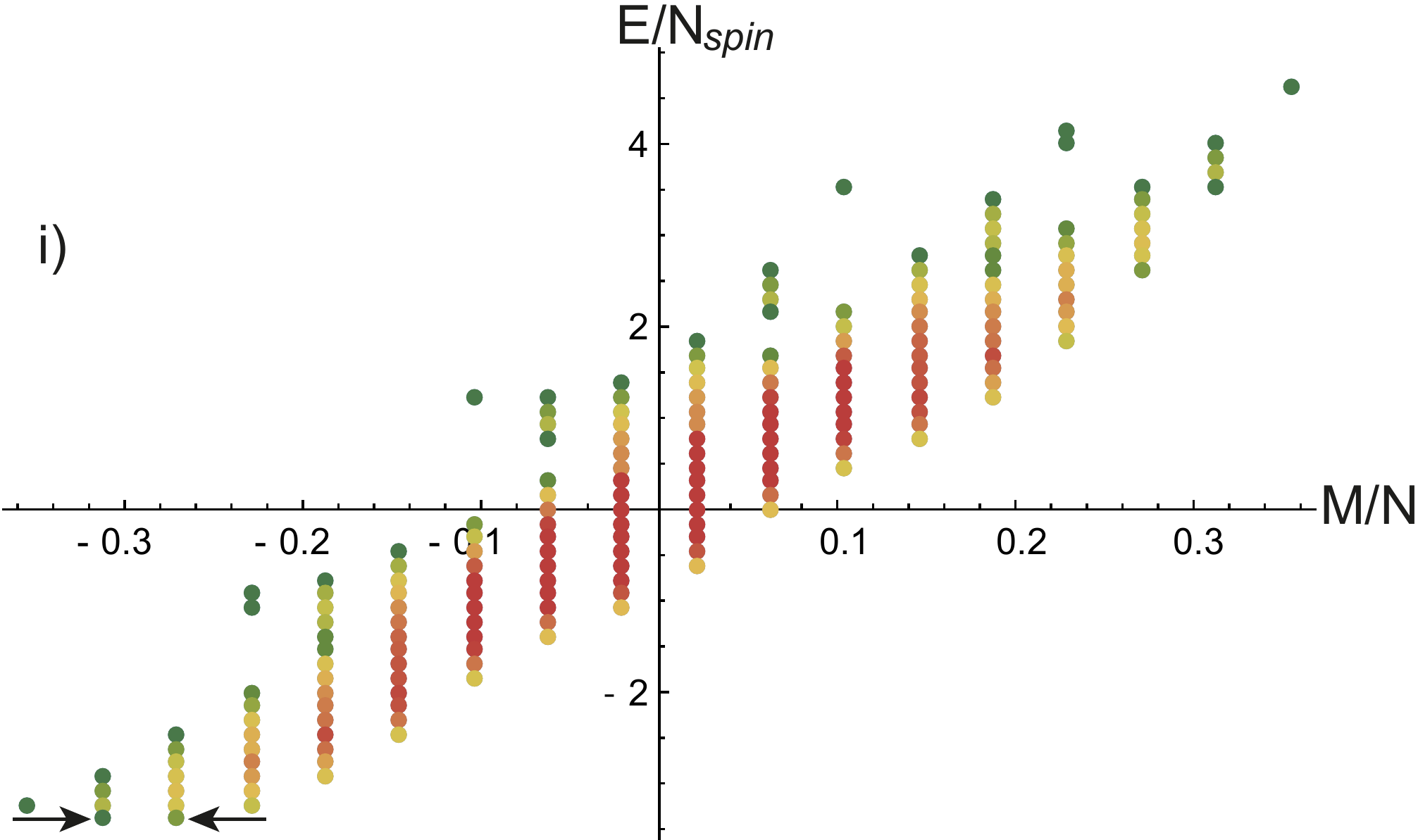}}
	\end{minipage}
	\vfill
	\begin{minipage}[h]{0.4\linewidth}
	\center{\includegraphics[width=1\linewidth]{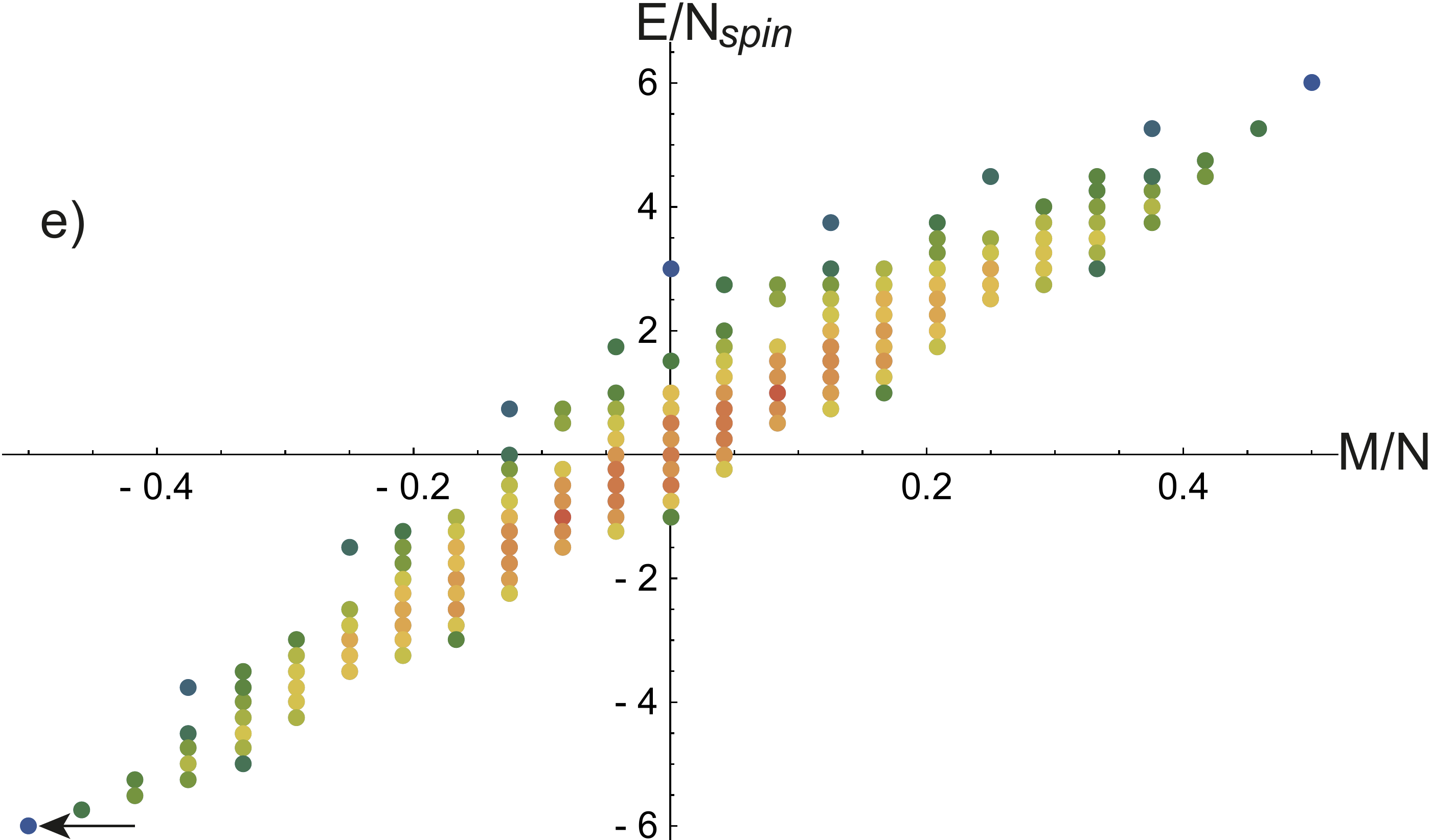}} 
	\end{minipage}
	\hfill $h/J=12$ \hfill
	\begin{minipage}[h]{0.4\linewidth}
	\center{\includegraphics[width=1\linewidth]{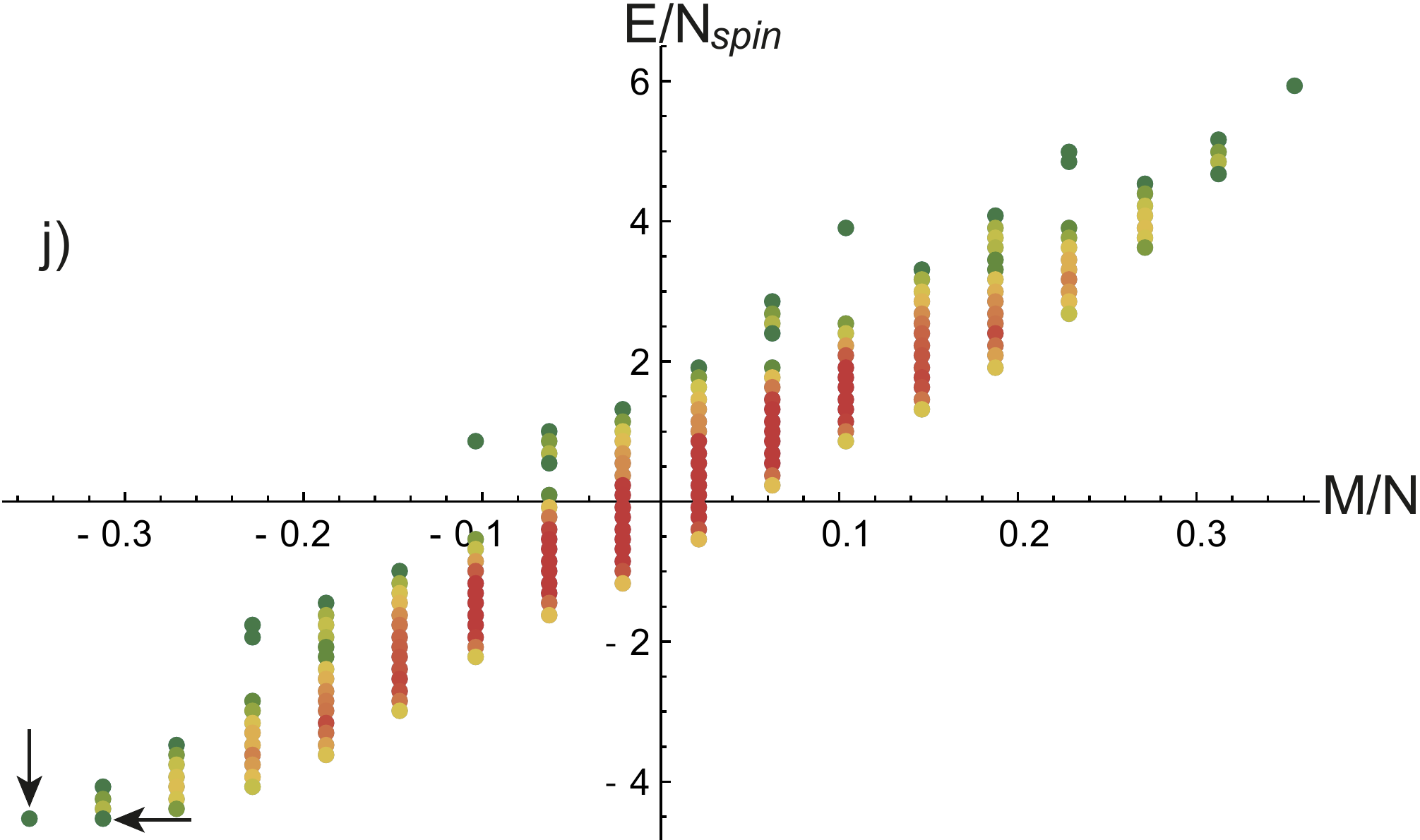}}
	\end{minipage}
	\caption{DOS for the AFM Ising model on the pyrochlore lattice for different external magnetic fields. The system size is L = 1 (N = 16), left column - pure system (a-e), right column - diluted system (f-j), dilution concentration x = 3/16.}
	\label{fig:explanation}
\end{figure*}
%%%%%%%%%%%%%%%%%%%%%%%%%%%%%%%%%%%%%%%%%%%%%%%%%%%%%%%%%%%%%%%%%%%%%%%%%%%%
In previous studies, authors investigated the origin of multiple magnetization 
plateaus by considering the local spin configuration
of triangles in the diluted antiferromagnetic Ising models on the pyrochlore, triangular and kagome lattices \cite{Soldatov, Peretyatko}. 
The qualitatively different behavior of the plateaus results from the 
competition between the exchange and Zeeman energies, which differ 
in the pyrochlore (Table I of Ref.  \cite{Peretyatko}), triangular and kagome lattices 
(Tables I and II of Ref. \cite{Soldatov}). 

In the present study, we show that when system is reaching the critical field, several
states with the different magnetization acquire the same 
total lowest energy value, which leads to the degeneracy of these states, 
i.e. large residual entropy peaks. The dilution effect leads to the new additional residual entropy peaks.
For example, in Fig.~\ref{fig:pyro_pure} we see two large peaks of the residual entropy in the absence of the magnetic field and at the field $h/J=6$. 
In Fig.~\ref{fig:pyro_dilute} one can notice peaks of the residual entropy at the same critical field values, and new peaks at $h/J = 3, 9, 12$ appear due to the dilution.

In Fig.~\ref{fig:explanation} we show the density of states of the pyrochlore lattice unit cell ($L$ = 1, $N$ = 16) for pure (left column) and diluted (right column, x = 3/16) systems in the magnetic fields $h/J = 0, 3, 6, 9, 12$.
Color scale on the top of the figure displays the logarithm of the degeneracy of states per spin, arrow marks pointing to lowest energy states.

In Fig.~\ref{fig:explanation}a and Fig.~\ref{fig:explanation}f we see five different lowest energy states for pure system and six lowest energy states for diluted system.
We can calculate ground state magnetization per site as averaging over all different ground state points as $M/N = ((0+ 1/3+1/6 -1/3 - 1/6)/5)/16 =0$ for pure system and $M/N = ((3/16+ 7/48+1/48 -1/48 - 7/48- 3/16)/6)/16 = 0$ for diluted system.
Also, we can obtain the residual entropy per spin of the pure system as $S/N_{spin} = ln(24+9+24+24+9)/16 = 0.281$ and $S/N_{spin} = ln(112+96+160+160+96+112)/13 = 0.508$ for diluted case.
The summation of the degeneracy of these states leads to a peak of the residual entropy with the zero value of the average ground state magnetization.
Even the small external magnetic field leads to abrupt change in the ground state magnetization and the residual entropy.

Figure~\ref{fig:explanation}b shows that with an addition of the magnetic field $h/J = 3$, ground state has a magnetization value $M/N = 1/3$ and residual entropy is $S/N_{spin} = ln(9)/16 = 0.137$ for pure system.
This state continues to be a ground state in all fields from $0<h/J<6$, which corresponds to the plateau of ground state magnetization and residual entropy, as shown in Fig.~\ref{fig:pyro_pure}. 
Energy states which also were lowest at $h/J = 0$, were increased due to the addition of Zeeman energy. 
In contrast, Fig.~\ref{fig:pyro_dilute} displays a peak of the residual entropy and a step in the ground state magnetization at $h/J = 3$  for diluted systems. 
This effect appears due to the summation of the two lowest energy states with different magnetizations $M/N = (-3/16 - 11/48)/2 = -0.208$, as can be seen in Fig.~\ref{fig:explanation}g.

When reaching the critical field $h/J =6$, we see another five different lowest energy states for the pure system in Fig.~\ref{fig:explanation}c, and two lowest energy states for the diluted system in Fig.~\ref{fig:explanation}h. 
The multiplicities of the degeneration of these states are summarized, which leads to the peak of residual entropy and magnetization step for the pure and diluted systems. 
Since there are five multiplicities of the ground states with different magnetizations for a pure system and the only two for diluted, the peak of the pure system will be much higher than in any diluted case.
If we further increase the field, the energies of these states increase due to their addition of Zeeman energy, and only one true ground state configuration remains for pure system (Fig.~\ref{fig:explanation}d,e). 
Residual entropy in pure system for all fields above the critical value $h/J = 6$ is equals to $ln(1)/N = 0$.

In contrast, in the diluted systems (Fig.~\ref{fig:explanation}i,j) we see the degeneracy of the ground states in two critical fields $h/J =9$, $h/J = 12$. 
As seen in Fig.~\ref{fig:pyro_dilute}, the peaks of the residual entropy and the steps of the ground state magnetization appear at these values of the field. A further increase of the field leads to such a strong dominance of Zeeman energy over the exchange energy that only one ground state with the maximum absolute value of the magnetization remains.
Thus, the residual entropy peaks and the ground state magnetization steps in the diluted pyrochlore lattices appear due to the discrete structure of the density of states, which transforms when an external magnetic field excites the system. 
%In contrast, в разбавленных системах (Рисунки 11 - I,J) мы видим выраждения основных состояний при двух критических полях $h/J =9$, $h/J = 12$, что приводит к пикам остаточной энтропии и скачкам намагниченности при этих значениях поля. 
%Дальнейшее увеличение поля приводит к настолько сильному преобладанию энергиии Зимана над обменной энергией, что остается только один граунд стейт с максимальным абсолютным значением намагниченности.  
%Таким образом, пики остаточной энтропии и скачки намагниченности в разбавленных системах решетки пирохлора происходят из за дискретной структуры дос которая перестраивается в зависимости от разбавления. При добавлении внешнего магнитного поля дос начинается наколнятся в одну из сторон и при критических значениях полей происходит вырождение граунд стейт. Значения критическх полей можно получить с помощью рассмотрения локальных спиновых конфигураций в треугольниках как это было сделано в работах \cite{Soldatov, Peretyatko}.

\section{Summary and Discussions}

We studied the diluted spin-ice model on the pyrochlore lattice 
when a magnetic field is applied in the [111] direction. 
In order to investigate the entropy, we use the WL Monte Carlo 
method~\cite{WL}, which directly calculates the energy DOS.  

We obtained the multiple magnetization plateaus for the AFM Ising model 
on the pyrochlore lattice with a magnetic field in the [111] direction, 
which is consistent with the previous calculation using the replica-exchange 
Monte Carlo simulation~\cite{Peretyatko}.  
We observed the stepwise decrease of the residual entropy, and 
large peaks at the crossover magnetic fields. 

We also observed the multiple magnetization plateaus 
for the AFM Ising model on the kagome lattice in a magnetic field. 
Although the pyrochlore lattice can be considered as the alternative 
layers of kagome and triangular lattices, and the kagome layers 
are the subject of interest when the magnetic field is applied 
in the [111] direction, the effect of dilution is different.  
The positions of the magnetization steps for the AFM Ising model 
on the kagome lattice are $h/J$ = 1, 2, 3, and 4, 
as shown in Fig.~\ref{fig:kagome_dilute}. 
Large peaks in the entropy are observed at the magnetic 
fields where the magnetization steps appear. 

We observed seven magnetization plateaus in the AFM Ising 
model on the triangular lattice with the magnetization steps 
at $h/J$ = 1, 2, 3, 4, 5, and 6.  
Although there is no macroscopic residual entropy 
for the pure triangular lattice, we found a residual entropy 
for the diluted model, which can be regarded 
as the residual entropy due to the dilution.  
We observed the large peaks in the entropy at the crossover 
magnetic fields for the triangular lattice. 

In the Appendix, we summarize theoretical results of 
the magnetization plateaus and the residual entropy 
for the pure AFM Ising model on the triangular, kagome, 
and pyrochlore lattices in magnetic fields.  
In the case of the pyrochlore lattice, 
the magnetic field is applied in the [111] direction.
As a byproduct, we give the exact numerical estimates of 
the magnetization, as given in Eq.~(\ref{Baxter_mag}), and entropy, 
as given in Eq.~(\ref{Baxter_entropy}), up to 16 digits 
at the crossover field $h_c/J=6$ 
for the pure AFM Ising model on the triangular lattice 
in a magnetic field based on the exact solution reported by 
Baxter~\cite{Baxter1}. 

We observed large residual entropy peaks for the diluted systems, 
and these peaks correspond to the multiple magnetization plateaus. 
Isakov {\it et al.}~\cite{Isakov04} reported such an entropy peak 
in the pure spin-ice model in a magnetic field, and queried 
the generality of this phenomenon for finite temperatures. 
In this paper we have shown that large entropy peaks do appear 
even for random systems, such as diluted systems. 
At the crossover magnetic fields, higher-magnetization states 
and lower-magnetization states are degenerate. 
The mixture of the local spin configuration of both states 
yield macroscopic large entropy. 
The change of the states comes from the competition 
between the exchange and Zeeman energies. Such competition 
is more complicated in the diluted model than in the pure 
model, which was analyzed  previously~\cite{Peretyatko}. 

The magnetic field and dilution offer a rich variety 
of effects in frustrated systems, 
and the application of the present study to other models 
of frustration is currently in progress.

\section*{Acknowledgment}

We thank Vitalii Kapitan for valuable discussions. 
The computer cluster of Far Eastern Federal University 
and the equipment of Shared Facility Center “Data Center of FEB RAS” (Khabarovsk) were used for computation.

This work was supported by a Grant-in-Aid for Scientific Research 
from the Japan Society for the Promotion of Science No. JP16K05480, by the Russian Ministry of 
Science and Education state contract no. 3.7383.2017/8.9, research project No. 18-32-00557 by RFBR and by a grant from the President of the Russian Federation for young scientists and graduate students, in accordance with the Program of Development Priority Direction “Strategic information technologies, including the creation of supercomputers and software development”, grant \#SP-4348.2018.5.

\appendix

\section{Theory of pure systems}

We review the theory of pure AFM Ising models 
in a magnetic field on the frustrated lattices in the Appendix. 
We start with the triangular lattice. 
The AFM Ising model on the 2D triangular lattice without a magnetic field 
was exactly solved by Wannier~\cite{Wannier}, and it was shown 
that at all the temperatures this system has no long-range order 
due to frustration. The residual entropy of the AFM Ising model 
on the triangular lattice without magnetic field was calculated 
to be 0.323066~\cite{Wannier}:
$$
  s_{h=0} = \frac{S_{h=0}}{N} = \frac{2}{\pi} \int_0^{\pi/3} \ln(2\cos \omega) 
      \ d\omega = 0.323066.
$$
When a magnetic field is applied, the system takes the "up-up-down" 
spin configuration in a basic triangle, which results in the 1/3 
magnetization plateau. 
This is the case of a weak magnetic field ($0 < h/J < 6$).  
For a strong magnetic field ($h/J > 6$), the magnetization 
jumps to the saturated value, $m = M/N =1$, 
and the jump becomes smoother with increasing temperature. 

As another example of the 2D frustrated lattice, the AFM 
Ising model on the kagome lattice without magnetic field 
was exactly solved by Kano and Naya~\cite{Kano}.
The residual entropy of the AFM Ising model 
on the kagome lattice was calculated to be 0.50183~\cite{Kano}:
\begin{eqnarray*}
  s_{h=0} &=& \frac{S_{h=0}}{N} = \frac{1}{24\pi^2} \int_0^{2\pi} \int_0^{2\pi} 
  \ln [21-4 (\cos \omega_1 + \\ &~& \cos \omega_2 + \cos(\omega_1+\omega_2))] 
  \ d\omega_1 d\omega_2 = 0.50183.
\end{eqnarray*}
When a magnetic field is applied, the system shows the same 
behavior of the magnetization plateau as in the triangular lattice.
The crossover magnetic field is $h_c/J = 4$ for the kagome lattice. 

The behavior of the residual entropy is interesting. In the weak magnetic 
field region ($0 < h/J < 6$) for the triangular lattice, 
all the spin configuration is determined once the "down" spin 
in one basic triangle is selected because of the closed packed structure 
of the triangular lattice, which means that the system has 
a three-fold degeneracy and there is no macroscopic degeneracy.  
On the contrary, for the kagome lattice, there is a macroscopic 
degeneracy for $0 < h/J < 4$ because of the loose packed structure. 
It is shown that the AFM Ising model of the kagome lattice 
in the 1/3 magnetization plateau 
region is equivalent to the dimer problem in the honeycomb 
lattice~\cite{Udagawa,Moessner,Isakov04}. 
The entropy is calculated to be one-third of the Wannier value of 
the residual entropy of the triangular lattice without magnetic field, 
that is, 0.107689 \cite{comment}. 

For a strong magnetic field region, i.e., $h/J>6$ for the triangular lattice 
and $h/J>4$ for the kagome lattice, all of the spins are fixed as "up", 
and there is no degeneracy. At the crossover magnetic field, $h_c$, 
all of the spin configurations of the "up-up-down" configuration or 
the "up-up-up" configuration in each triangle have the same energy. 
Thus, there appears to be a large macroscopic degeneracy, which is associated 
with the large peak in the entropy~\cite{Isakov04}.  

The AFM Ising model on the triangular lattice at $h/J=6$ 
is equivalent to the hard hexagon model with the activity $z = 1$. 
The hard hexagon model~\cite{Baxter1,Baxter2} 
is a 2D lattice gas model, where particles are allowed 
to be on the vertices of a triangular lattice but no two particles 
may be adjacent. The condition that $z=1$ is applied when the chemical 
potential is zero. 
Metcalf and Yang~\cite{Metcalf} performed approximate numerical studies using 
the transfer matrix, and they obtained the average magnetization 
and residual entropy per spin as $m=0.6751$ and $s=0.3333$, 
respectively. They conjectured that the exact value of 
the entropy is $1/3$.  Baxter and Tsang~\cite{Baxter3} extended 
numerical studies using the corner-transfer matrix (CTM) method, 
and they obtained a more accurate numerical estimate, namely 
$0.333 242 721 976$, which contradicts the conjecture.
Finally, Baxter exactly solved the hard hexagon 
model, and confirmed their assertion~\cite{Baxter1,Baxter2}. 
The partition function $\kappa$ and the activity $z$ are expressed 
by infinite products of a variable $x$, 
but the explicit expressions of the magnetization and the entropy 
were not given in the literature. 
Thus, here, we show the exact numerical estimates 
of the magnetization and the entropy. 
By solving the infinite products numerically, 
we obtain 
\begin{eqnarray}
 m = 0.6751341572050237
\label{Baxter_mag}
\end{eqnarray}
and
\begin{eqnarray}
 s = 0.3332427219761819
\label{Baxter_entropy}
\end{eqnarray}
up to 16 digits. 
The CTM estimate of the entropy~\cite{Baxter3} 
is shown to have been correct up to 12 digits.

The AFM Ising model on the kagome lattice at $h/J=4$ 
is equivalent to the monomer-dimer mixture with the activities 
$z_2/z_1^2=1$, where $z_1$ and $z_2$ are the activity of 
monomers and that of dimers, respectively, 
in the honeycomb lattice \cite{Isakov04}. 
In the dimer problem, which is equivalent to the 
kagome lattice AFM Ising model for $0<h/J<4$, all of the vertices in 
the honeycomb lattice consist of dimers. 
If there are vertices that are not part of any dimer, 
they are called monomers. The weights of monomers 
and dimers are the same for the problem of $h_c$ 
of the kagome lattice, which yields $z_2/z_1^2=1$. 
The Bethe approximation of this model was discussed 
by Nagle \cite{Nagle2}. 
The grand partition function was given as a function 
of the coordination number of the lattice $q$. 
When we use $q=3$ for the honeycomb lattice, 
we obtain the magnetization $m=3/5$ and 
the entropy $s = (1/3) \ln (16/5) = 0.388$~\cite{Isakov04}.

%%%%%%%%%%%%%%%%%%%%%%%%%%%%%%%%%%%%%%%%%%%%%%%%%%%%%%%%%%%%%%%%%%%%%%%%%%%%
\begin{table}[t]
\caption{
Ground-state magnetization and residual entropy per spin as a function of $h$ 
for the AFM Ising models on the triangular, kagome and pyrochlore lattices, 
where $h_c = 6J$ for the triangular and pyrochlore lattices and 
$h_c = 4J$ for the kagome lattice.
}
\begin{center}
\begin{tabular}{lllll}
\hline
\hline
lattice \quad & $h=0$ \quad & $0<h<h_c$ \quad & $h=h_c$ \quad & $h>h_c$ \quad \\
\hline
$m$ (triangular) & \ 0       & \ 1/3         & \ $m_1$        & \ 1        \\
$s$ (triangular) & \ $s_1$ & \ $\ln(3)/N$    & \ $s_4$      & \ $\ln(1)/N$ \\
\hline
$m$ (kagome)     & \ 0       & \ 1/3         & \ $m_2$     & \ 1        \\
$s$ (kagome)     & \ $s_2$ & \ $s_1/3$ & \ $s_5$ & \ $\ln(1)/N$ \\
\hline
$m$ (pyrochlore) & \ 0       & \ 1/3         & \ $(m_2+1)/4$        & \ 1/2      \\
$s$ (pyrochlore) & \ $s_3$ & \ $s_1/4$ & \ $(3/4)s_5$      & \ $\ln(1)/N$ \\
\hline
\hline
\end{tabular}
\end{center}
\begin{flushleft}
$s_1=0.323066$ (exact) \cite{Wannier}. \\
$s_2=0.50183$ (exact) \cite{Kano}. \\
$s_3=0.20501$ \cite{Nagle}, which is close to Pauling's estimate 
$(1/2)\ln (3/2)=0.20273$ \cite{Pauling}. \\
$m_1 = 0.6751341572050237$ (exact) \cite{Baxter1,Baxter2,Otsuka},
and a previous numerical estimate was given in Ref.~\cite{Metcalf}. \\
$s_4$ = 0.3332427219761819 (exact) \cite{Baxter1,Baxter2,Otsuka}, 
and previous numerical estimates were given in Refs.~\cite{Metcalf,Baxter3}. \\
$m_2=3/5$ (Bethe approximation \cite{Nagle2,Isakov04}). \\
$s_5=(1/3)\ln(16/5)=0.388$ (Bethe approximation \cite{Nagle2,Isakov04}). \\
\end{flushleft}
\end{table}
%%%%%%%%%%%%%%%%%%%%%%%%%%%%%%%%%%%%%%%%%%%%%%%%%%%%%%%%%%%%%%%%%%%%%%%%%%%%

The pyrochlore lattice can be regarded as an alternating sequence 
of kagome and triangular layers that become effectively decoupled 
by a magnetic field oriented along the [111] direction. 
The behavior of the spins in the kagome layers is of significant interest, 
and is sometimes referred to as the "kagome-ice" problem. 
The problem of the three-dimensional pyrochlore lattice in the 
[111] magnetic field is essentially the same as 
that of the 2D kagome lattice in the magnetic field. 
Because the number of contributed spins is different, 
the entropy per spin of the pyrochlore lattice 
is related to that of the kagome lattice as 
\begin{eqnarray*}
 s_{{\rm pyrochlore}} = (3/4)s_{{\rm kagome}}.
\end{eqnarray*}
The relation between the magnetization of the pyrochlore lattice problem 
and that of the kagome lattice problem is 
\begin{eqnarray*}
 m_{{\rm pyrochlore}} = (m_{{\rm kagome}} +1)/4. 
\end{eqnarray*}
The magnetization plateau for the weak magnetic field is 1/3, 
whereas the saturated value of the magnetization becomes 1/2 
for the pyrochlore lattice in the magnetic field 
along the [111] direction. 
The equivalence of the pyrochlore lattice and the kagome 
lattice is only for the systems with magnetic field.  The residual 
entropy of the pyrochlore lattice without magnetic 
field was calculated by Nagle~\cite{Nagle} as 0.20501, 
which is close to Pauling's estimate 
$(1/2)\ln (3/2)=0.20273$ \cite{Pauling}.

We tabulate the theoretical results of the ground-state magnetization 
and residual entropy as a function of $h$ 
for the AFM Ising models on the triangular, kagome and pyrochlore 
lattices in Table II, 
where $h_c = 6J$ for the triangular and pyrochlore lattices and 
$h_c = 4J$ for the kagome lattice.

We also show the plots of the magnetization and residual entropy 
as a function of $h$ 
for the AFM Ising models on the pyrochlore, kagome and triangular lattices 
in Figs.~\ref{fig:pyro_pure}, \ref{fig:kagome_pure}, 
and \ref{fig:tri_pure}, respectively. 
We plot the numerical data using the WL method 
in the figure. 
We can see that the WL method agrees well with 
the theoretical values.


\begin{thebibliography}{99}
\bibitem{comment}
The expression of the residual entropy given in 
Ref.~\cite{Udagawa} is different from that of 
Wannier~\cite{Wannier}. However, we can show that they are equivalent 
using the integration by substitution.
\end{thebibliography}
\end{document}